\documentclass[preprint2]{aastex}

\usepackage{epsfig}
\usepackage{longtable}
\usepackage{amssymb}
\usepackage{amsmath}
\usepackage{graphics}
\usepackage{times}
\usepackage{epstopdf}
\usepackage{latexsym}
\usepackage{amsmath}
\usepackage{amsfonts}
\usepackage{amssymb}
\usepackage{multicol}
\usepackage{array}
\usepackage{multirow}
\usepackage{subfigure}

\graphicspath{{./}{figures/}}

\usepackage{fancyhdr}   
\usepackage{lipsum}     

\shorttitle{Deep H$\alpha$ survey of the Coma cluster}
\shortauthors{Kay et al.}

\begin{document}

\title{Deep H$\alpha$ survey of the Coma cluster: The Catalog}

\author{Sarah E. Kay}
\affil{Department of Physics and Astronomy, University of Wyoming}
\email{seftekha@uwyo.edu}

\author{Ehsan Kourkchi}
\affil{Institute for Astronomy, University of Hawai‘i, Honolulu, HI, USA}
\email{ehsan20@hawaii.edu}

\author{A. Molaeinezhad and H. G. Khosroshahi}
\affil{School of Astronomy, Institute for Research in Fundamental Sciences (IPM), 
PO Box 19395-5531, Tehran, Iran}

\author{M. Mouhcine}
\affil{Observatoire Astronomique de Strasbourg, 
11 rue de l'Universit\'e, 67000 Strasbourg, France}

\author{P. A. James and D. Carter}
\affil{Astrophysics Research Institute, Liverpool John Moores University, 
Twelve Quays House, Egerton Wharf, Birkenhead CH41 1LD, UK}

\begin{abstract}
We present a deep wide-field narrow-band imaging survey of the local rich and dynamically relaxed Coma cluster of galaxies, carried out with the Wide Field Camera at the Isaac Newton Telescope. The survey covers a region of about 2.5 sq. deg. extending from the core of the cluster out to the infall region over the south-west quadrant of the Coma cluster. The $R$ (6380~\AA) and $[$S$\scriptstyle\rm II$$]$ (6725~\AA) filters of WFC/INT were used to derive the H$\alpha$+[N{\sc ii}] fluxes and equivalent widths of cluster galaxies distributed over a wide range of environmental conditions. The depth of our imaging observations allows us to measure reliably those properties well down into the dwarf regime in the Coma cluster for the first time. We have detected 124 H$\alpha$ emitting sources with spectroscopically-determined membership, 96 of which have not been detected previously. In this paper, we report on the data analysis process and the methodology we used to measure reliable H$\alpha$ properties, and present the measurement catalogue. 
\end{abstract}


\section{Introduction}
\label{Sec:introduction}

\subsection{Background}

Some of the most fundamental questions in galaxy evolution concern the processes of ongoing star formation, and the internal and external influences on a galaxy which contribute to either the enhancement or suppression of star formation activity.  Many techniques involving all parts of the electromagnetic spectrum now contribute to our understanding of star formation, but one of the oldest and best calibrated methods, imaging in the Balmer H$\alpha$ line from excited and ionized hydrogen gas, remains one of the most powerful and sensitive methods. Early work by Kennicutt and collaborators \citep{ken83} and subsequent developments have established that current massive star formation activity is accurately traced by the integrated H$\alpha$ line emission. The application of H$\alpha$ techniques to the star formation properties of galaxies in the local Universe has been fully reviewed by Kennicutt (1998).

An area of particular interest concerns the effect of environment (cluster, group or field) on star formation histories of galaxies. It is well established that the morphology of a galaxy correlates with its environment, i.e., the so-called morphology-density relation \citep{dre80}. An aspect of this correlation is the general trend for star formation activity to be suppressed in dense environments relative to the field \citep{lew02}. The nature of the driver(s) of the observed predominance of spheroid-dominated morphologies and a paucity of recent star formation in local clusters is still debated. H$\alpha$-based studies of galaxies in the cluster environment have played a central role throughout this debate, starting by the early attempt of \citet{ken83} to compare the properties of the Virgo cluster spiral galaxies to their field counterparts. Since then a number of local galaxy clusters have been targeted, e.g., Virgo \citep{koo98,gav02,bos02}, Coma  \citep{ken84,gav91,gav06,igl02}, Abell 1367  \citep{mos88,sak02,gav03,cor06,kri11}, and a number of local Abell clusters \citep{mos00,tho08,ced09,bre10}. For a full discussion of the history of H$\alpha$ studies of nearby cluster galaxies and their results, see \citet{bos06}.  Apart from the Virgo cluster, for which the H$\alpha$-selected samples cover extended ranges of galaxy morphologies and masses \citep{bos02,koo06}, the published samples of H$\alpha$-selected cluster galaxies are sampling primarily bright spiral galaxies. 

Coma is the nearest rich cluster of galaxies, and harbors regions of widely varying environmental conditions. As such, it provides a key reference to study the role of the environment on the evolution of galaxies.  It has been targeted previously in H$\alpha$ (e.g. Whittle \& Moss 2000, 2005; Iglesias-P\'aramo et al. 2002; Gavazzi et al. 1998). Due to its redshift and its large angular extent, deep and wide-field surveys of this cluster have been limited. Previous surveys of the cluster have been able to detect only H$\alpha$ emission from relatively bright galaxies, i.e. $M_B$ $\sim -18$.  Early studies of the H$\alpha$ properties of the Coma cluster galaxies suffer from a number of biases. For example, the samples of the Coma cluster H$\alpha$ emitters of Kennicutt et al. (1984) and Gavazzi et al. (1991, 1998) were selected on the basis of their optical properties. On the other hand, even though the objective-prism surveys by Moss et al. (1988, 1998) and Moss $\&$ Whittle (2000) were H$\alpha$ selected, they were limited to the bright end of galaxy luminosity function and missed low surface brightness extended H$\alpha$ emission. The most directly comparable survey to the present work is that conducted by Iglesias-P\'aramo et al. (2002) who undertook a narrow-band imaging survey of the central one square degree of the Coma cluster.

Recently, \citet{yos08} and \citet{yag10} have presented some very deep imaging of the core of
the Coma cluster, and identified a number of extended structures which they propose are recent and ongoing
stripping events.

\begin{figure}[t]
\begin{center}
\includegraphics[width=0.47\textwidth,angle=0]{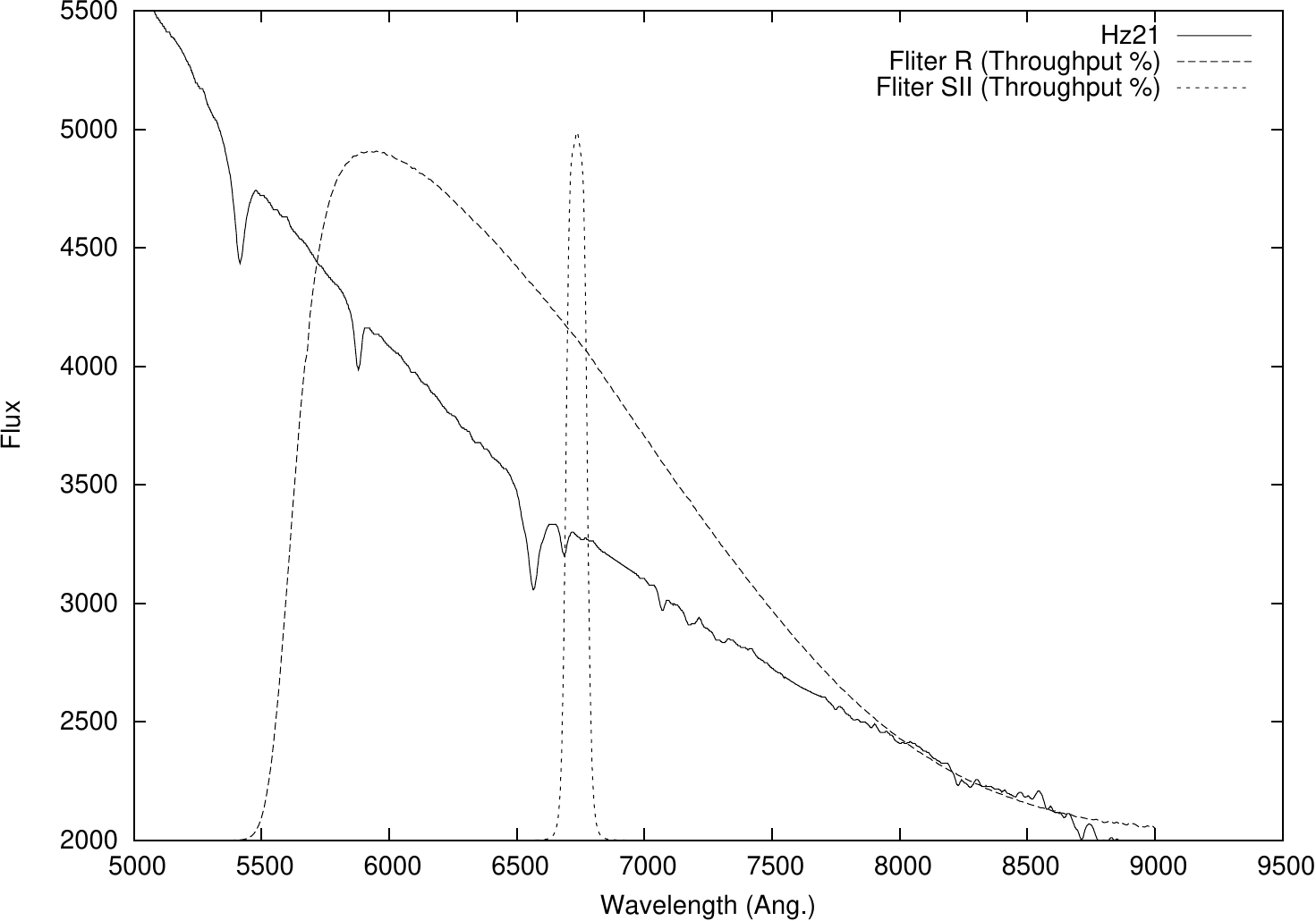}
\caption{The filter throughputs ($R$ and $[$S$\scriptstyle\rm II$$]$) and the spectrum of the observed standard spectrophotometric star, Hz21.}
\label{fig:hz210}
\end{center}
\end{figure}

\begin{figure}[t]
\begin{center}
\hspace{-0.75cm}
\includegraphics[width=0.50\textwidth]{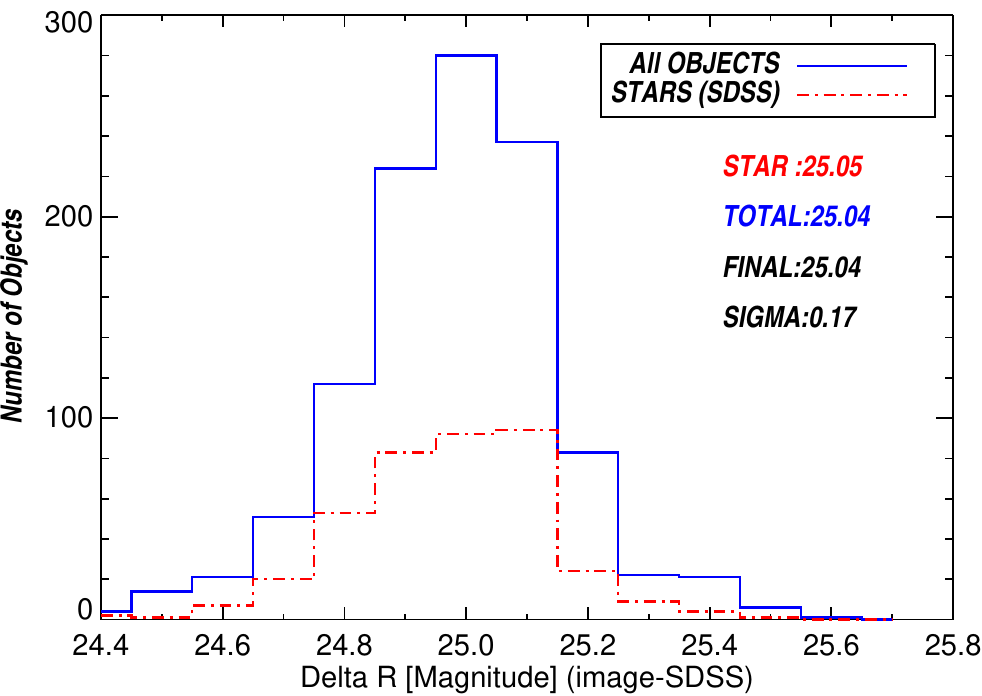}
\caption{The distribution of the magnitude shift for the point sources in the {\it Coma-core1} pointing (see Tables \ref{lisint} and \ref{lispo}). Delta $R$-magnitudes is the difference between the SDSS magnitudes and  that from {\it SExtractor} on our $R$-band images. About 50 percent of the point sources, which were found with {\it starfind}, were not confirmed by SDSS as stars. The blue solid histogram represents all point sources and the red dashed line shows the same for objects confirmed by SDSS as stars. The median of the Delta $R$-magnitudes is calculated to find the magnitude zero point. The final zero point of our images in the $R$-band is 25.1$\pm$0.2 mag.
}
\label{Rstack}
\end{center}
\end{figure}

\subsection{The Coma Cluster Survey}
\label{survey}

As clusters of galaxies form, they accumulate the products of earlier hierarchical assembly, from individual galaxies to dense groups. Because galaxies enter clusters with rich histories of their own, it has been difficult to answer one of the basic questions of cluster formation: did clusters largely inherit their galaxy populations, or did they play a central role in shaping them? To identify the sites where the environmental impact on galaxies is effective, morphologies and energy distributions of galaxies covering a wide range of environments are required. Much of the effort dedicated to  quantify the environment influence on galaxies was generally limited to luminous galaxies. Dwarf galaxies--the most numerous and the most sensitive to environmental stimuli owing to their less strong gravity-- have long been out of reach of studies beyond the immediate local Universe. 

To understand the environmental-driven processes involved in shaping galaxy properties, it is essential to conduct studies of both massive {\it and} dwarf galaxies, and to {\it differentiate} between the relative influence of the local and the global galaxy environment on their properties. To constrain the relative importance of environmental processes, e.g. major mergers, high speed impulsive encounters, ram-pressure effects, on galaxy evolution, and to provide the long awaited local ``rich cluster'' benchmark for comparison with surveys of less dense and relaxed local galaxy clusters, and with distant cluster and field surveys, the Coma cluster is the focus of a intensive and coordinated multi-wavelength observational campaign. The panchromatic data-sets of the Coma cluster galaxies include deep XMM \citep{bri01}, {\it Chandra} \citep{vik01}, GALEX \citep{ham10}, Hubble Space Telescope \citep{car08}, {\it Spitzer} \citep{{jen07},{bai06}}, Herschel (PI: C.J. Simpson), and VLA \citep{mil09} observations, as well as deep ground-based optical and near-infrared imaging and optical spectroscopy. 

As part of this broad effort to characterize comprehensively the environmental impact on galaxies, and to overcome the limitations of previous narrow-band surveys, we are conducting a narrow-band imaging survey of the Coma cluster. The survey is designed to detect H$\alpha$ emission from all cluster star-forming galaxies with star formation rates down to a few times ${\rm 10^{-3} M_{\odot}\,yr^{-1}}$, similar to the rates measured for the Local Group dwarf galaxies. The area surveyed so far is about 2.5 square degrees extending from the core of the cluster out to the infall region to the south-west. The projected local galaxy densities change by a factor of about 100 over the targeted area. At the infall region, the survey is probing galaxies that have never passed through the core of the cluster, and are entering the cluster environment for the first time. The survey is then providing a unique H$\alpha$-selected sample of cluster galaxies for comprehensive statistical analysis of star formation activity in galaxies of rich relaxed clusters, and comparative studies with less dense clusters, the Virgo cluster in particular, and field galaxy samples. 

Future contributions in this series will use the catalogue and the images presented here, combined with the multi-wavelength data-sets we have in hand when necessary, to characterize the star formation activity in cluster galaxies, e.g. (i) global star formation properties, H${\alpha}$ luminosity function, star formation activity as a function of galaxy morphologies, stellar populations and environment (ii) spatial distribution of star formation activity within cluster galaxies (smooth vs. circum-nuclear vs. compact) and its dependence on environment and galaxy properties, (iii) the importance and the extent of preprocessing in the group environment, (iv) the relative importance of obscured and unobscured star formation in galaxies and its dependence on environment and (v) comparison of H$\alpha$ emission line properties to other star formation indicators sensitive to different timescales, e.g. H$\alpha$ vs. ultraviolet emission which is sensitive to star formation activity over a few 10$^{8}$ yr, in order to trace the recent star formation histories of cluster galaxies. This paper is organized as follows: The observations, properties of the data and comments on the observations are described in \S\ref{Sec:Observation}, while the reduction process and preparation of the data for calibration is described in \S\ref{Sec:Reduction}. Photometric and astrometric calibration and the procedure of continuum subtraction in order to obtain the properties of the detected H$\alpha$ emitters is explained in \S\ref{Sec:Photometry}. A short summary is given in \S\ref{Summary}.

\begin{figure*}
\begin{center}
\subfigure
{
\includegraphics[width=3in]{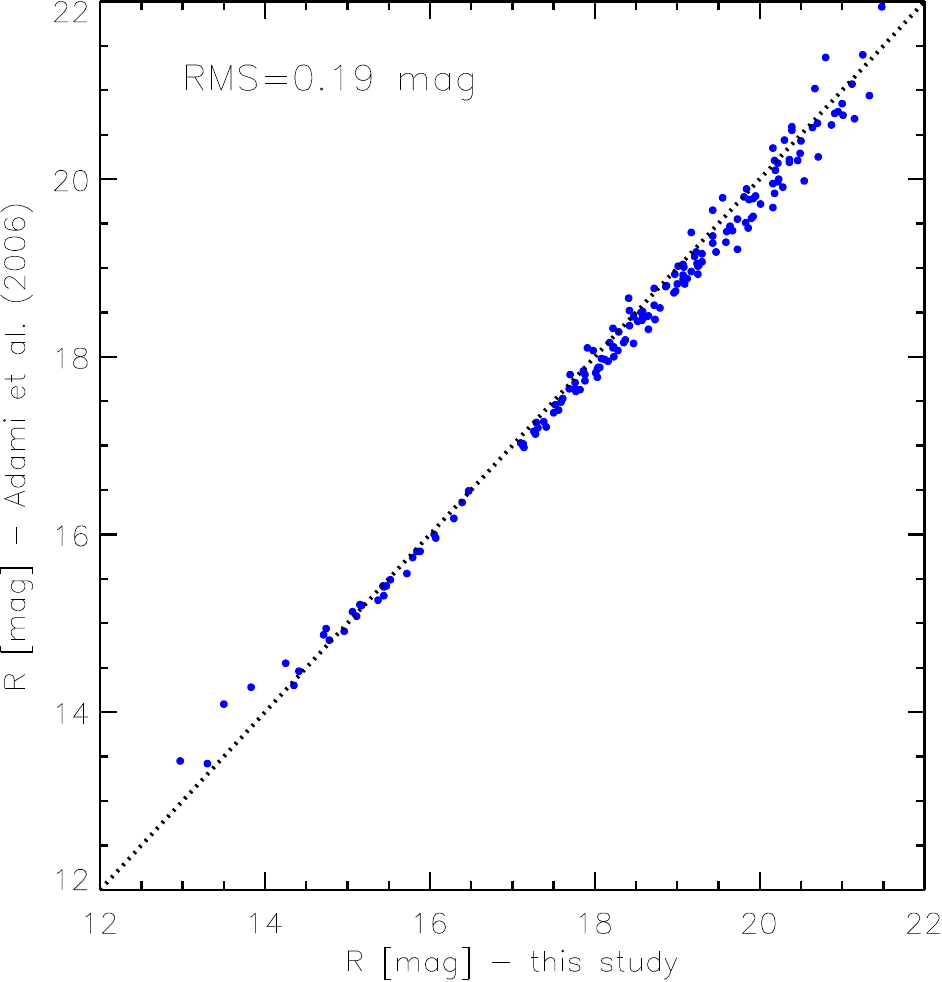}
}
\hspace{-0.1in}
\subfigure
{
\includegraphics[width=3in]{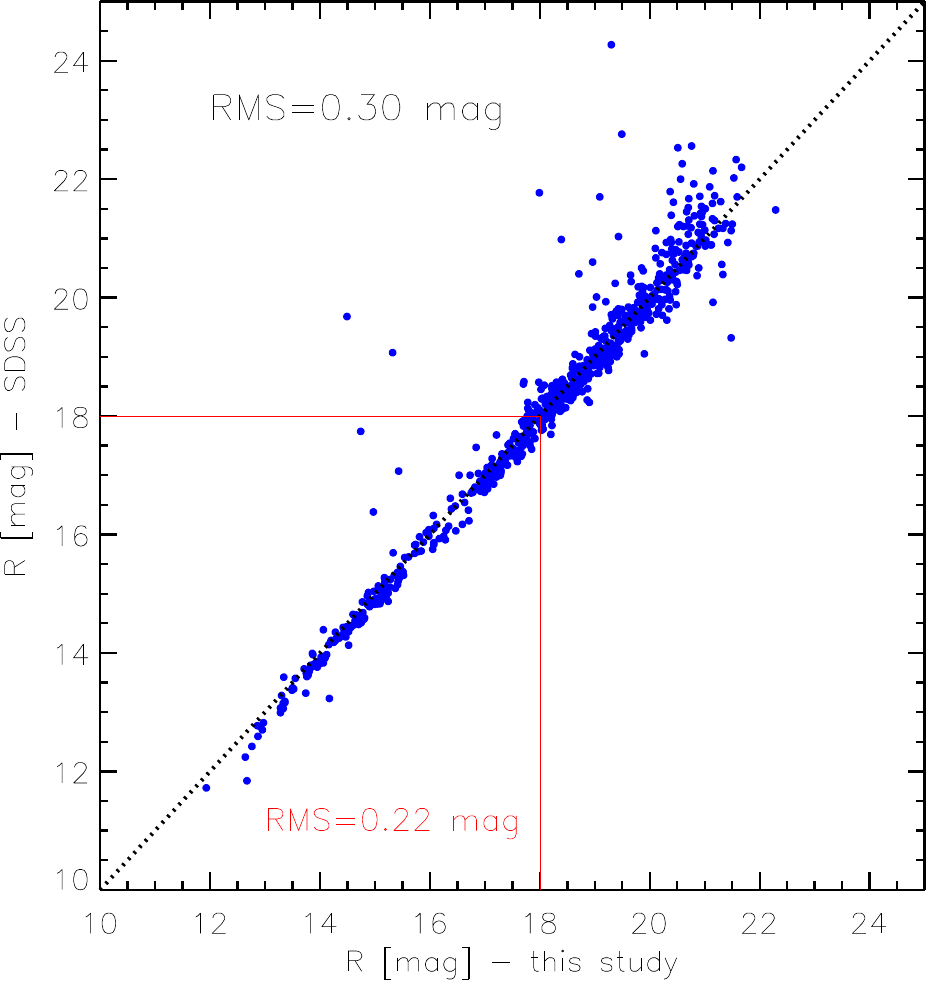}
}
\end{center}
\caption{Left: Comparison of the $R$ magnitudes from Adami et al. (2006) with those from this study. The measurements agree within the RMS scatter of 0.19 mag. Right: The SDSS derived $R$ magnitudes versus our measured $R$ magnitudes for common galaxies. The SDSS completeness limit in the $R$-band is $\sim$18 mag. For galaxies with $R \leq 18$ mag, the RMS scatter is 0.22 mag. Relaxing the luminosity cut, the RMS scatter is 0.30 mag.}
\label{fig:rcompare}
\end{figure*}

\section{Observations}
\label{Sec:Observation}

\begin{figure}[h]
\begin{center}
    \includegraphics[width=3.1in]{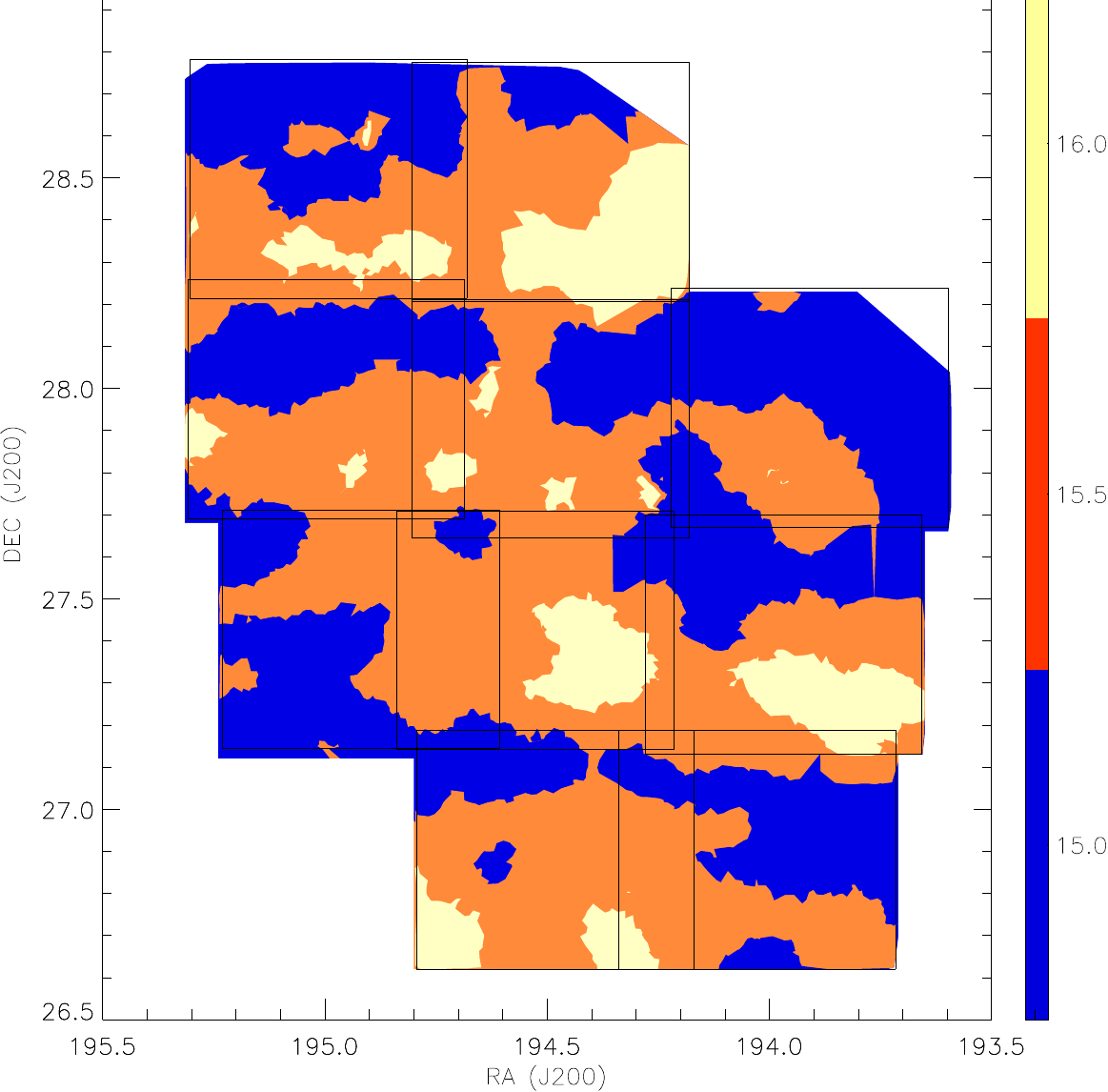}

\caption{The smoothed map of the flux scaling ratio (SR) variation across the observed field. The H$\alpha$+[N{\sc ii}] flux and EW of each object are calculated using the scaling ratio for the corresponding location.}
\label{ratiomap}
\end{center}
\end{figure}
\begin{figure}[h]
\begin{center}
\includegraphics[width=3in]{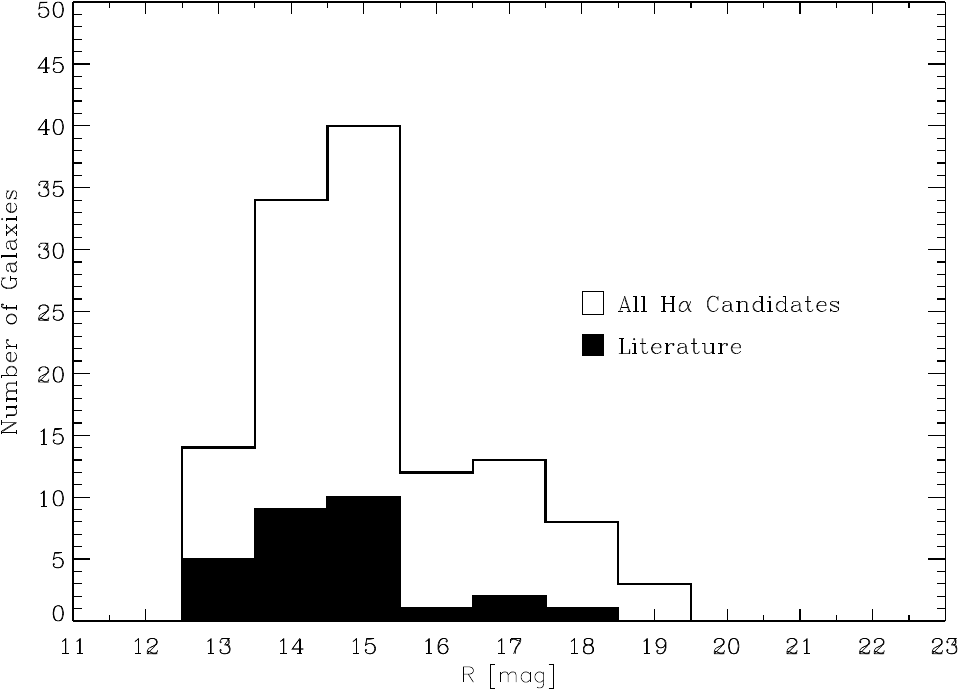}
\caption{The histogram of the $R$-band magnitudes of H$\alpha$ emitting members of the Coma cluster reported previously in the literature (black solid regions) and for those detected and confirmed Coma members in this survey (open histogram).
}
\label{Rhisto}
\end{center}
\end{figure}

The imaging was carried out with $R$ and $[$S$\scriptstyle\rm II$$]$ filters during the seven nights 
28$^{th}$ March -- 3$^{rd}$ April 2009 using the Wide Field Camera (WFC) mounted at the prime focus 
of the 2.5m Isaac Newton Telescope (INT) at Roque de Los Muchachos Observatory on La Palma. 
WFC consists of 4 thinned AR coated EEV CCDs with a size of $2048 \times 4096$ pixels and a pixel 
size of 13.5 micron (i.e. 0.333 arcsec/pixel), giving a total field of view about $34\arcmin \times 34\arcmin\ $. Given the arrangement of the detectors, a square area of about $11\arcmin \times 11\arcmin$ is lost at one corner  of the field.

Given the width of the $[$S$\scriptstyle\rm II$$]$ filter, with a central wavelength of 6725 \AA~ and pass-band width of 80 \AA, the [N{\sc ii}] 6548, 6584 \AA~ lines are also included in the high transmittance pass-band of this filter. Hereafter, the H$\alpha$ flux and equivalent width means H$\alpha$+[N{\sc ii}] flux and equivalent width, respectively. The $R$ band continuum observations were taken using a Harris $R$ filter with central wavelength of 6380 \AA~ and pass-band width of 1520 \AA. This broad $R$-band filter is used for continuum subtraction. 

The main camera and filter characteristics were obtained from the instrument information webpage, updated 
to the time of our observations. The North-East corner of the field of view suffers from vignetting. 
The exposure times were 300 sec and 1200 sec in the $R$-band and the $[$S$\scriptstyle\rm II$$]$-band, 
respectively. Table 1 gives the total integration time of each pointing in both pass-bands. Several zero-exposure 
frames and twilight sky exposures were taken at the beginning and end of all nights to determine the bias and 
flat field structure of each CCD.

Atmospheric turbulence produced an average PSF in the range $1.2 \lesssim $FWHM$ \lesssim 2.0$ arcsec in both $R$ and $[$S$\scriptstyle\rm II$$]$ filters, with small fluctuations across the field. Standard photometric star fields were observed during each night at different airmasses in order to correct the final images for atmospheric extinction and to determine the photometric zero point magnitude for each filter. Changes in the sky transparency during each night were monitored by regular $R$-band observations of standard stars selected from the list of Landolt (1992). The photometric calibration strategy for the narrow-band images (i.e. in the $[$S$\scriptstyle\rm II$$]$ band) was to observe at least one spectrophotometric standard star, in both filters at the beginning and end of each night. We used the Hz21 star from the ING list of standard stars. The throughput curves of the used filters and the spectrum of the observed spectrophotometric standard star (i.e. Hz21) are illustrated in Figure \ref{fig:hz210}.

\begin{table}
\renewcommand{\arraystretch}{0.99}\renewcommand{\tabcolsep}{0.12cm}
\caption{The list of pointings, exposure times and the dates of our observations with the WFC on the INT}
\label{lispo}
{\scriptsize
\begin{center}
\begin{tabular}{c c c c c c }
\tableline 
Date   &   Field  &   R.A.{*} &   Dec{*} & 
  EXP. (sec) & Filter \\
\tableline 
28$^{th}$ MAR. 2009 & F1 (SW) & 12:58:33 & 27:25:30 & [1200],300 & $[$S$\scriptstyle\rm II$$]$,R \\
    & F2 (SW) & 12:56:19 & 27:24:55 & [1200],300 &  $[$S$\scriptstyle\rm II$$]$,R \\
 &   &  &  & & \\
29$^{th}$ MAR. 2009 & F1 (SE) & 13:00:07 & 27:25:41 & [1200],300 & $[$S$\scriptstyle\rm II$$]$,R \\
   & F2 (SW) & 12:56:19 & 27:24:55 & [1200],300 & $[$S$\scriptstyle\rm II$$]$,R \\
 &   &  &  & & \\
30$^{th}$ MAR. 2009 & F3 (SW) & 12:56:33 & 26:54:14 & [1200],300 & $[$S$\scriptstyle\rm II$$]$,R \\
     & F4 (SW) & 12:58:22 & 26:54:14 & [1200],300 & $[$S$\scriptstyle\rm II$$]$,R \\
 &   &  &  & & \\
31$^{st}$ MAR. 2009 & F4 (SW) & 12:58:22 & 26:54:14 & [1200],300 & $[$S$\scriptstyle\rm II$$]$,R \\
     & CORE 1 & 13:00:26 & 27:58:25 & [1200],300 & $[$S$\scriptstyle\rm II$$]$,R \\
 &   &  &  & & \\
1$^{st}$ APR. 2009 & CORE 2 & 12:58:25 & 27:55:38 & [1200],300 & $[$S$\scriptstyle\rm II$$]$,R \\
     & CORE 4 & 13:00:25 & 28:29:49 & [1200],300 & $[$S$\scriptstyle\rm II$$]$,R \\
 &   &  &  & & \\
2$^{nd}$ APR. 2009 & CORE 2 & 12:58:25 & 27:55:38 & [1200],300 & $[$S$\scriptstyle\rm II$$]$,R \\
     & CORE 3 &  12:56:05 & 27:57:17 & [1200],300 & $[$S$\scriptstyle\rm II$$]$,R \\
 &   &  &  & & \\
3$^{rd}$ APR. 2009 & CORE 5 & 12:58:25 & 28:29:25 &  [1200],300 & $[$S$\scriptstyle\rm II$$]$,R \\[-0.2cm]
&   &  &  & & \\
\tableline 
\end{tabular}
\end{center}
}
{\footnotesize * R.A. and Dec of the pointings are in J2000 system.}
\end{table}

\section{Data Reduction}
\label{Sec:Reduction}

\subsection{Pre-reduction}
A large part of the data reduction was performed using {\it IRAF} standard packages. For bias subtraction, we used the mean value of the overscan region of each CCD as the bias value. This value was subtracted from the corresponding trimmed CCD image. We used the {\it IRAF}  task {\it ccdproc} for trimming the images. 

The ING group\footnote{INT Home Page: http://www.ing.iac.es} reported linearity departures of $\sim$2\% starting at $\sim$50,000 ADUs. CASU INT linearity coefficients\footnote{CASU INT Wide Field Survey Homepage: \\ http://www.ast.cam.ac.uk/$\sim$wfcsur/index.php} were applied to correct the nonlinearity, which are known to be very stable over the years, and are precise to 0.2\% in the 0-50 kilo-count range. We applied the latest linearity equations which were calculated by ING in 2008.

For flat-fielding, we first generated super-flat images obtained from the median average of all twilight exposures. The flat-fielding was performed by dividing all images (in $R$ and $[$S$\scriptstyle\rm II$$]$ filters) by their corresponding superskyflats. 

 For cosmic ray removal, we first used the {\it cosmicrays} task in {\it IRAF}. Since this task did not remove all cosmic rays, we used {\it SExtractor} \citep{ber96} to detect and mask those remaining. We used an IDL routine in order to include satellite tracks, diffracted and scattered light from bright objects (e.g. bright foreground stars and both cD galaxies at the center of the Coma cluster), saturated regions and other anomalies, in the final mask of each individual science frame. Before the co-addition process, the background gradient in some science frames was estimated and removed using the {\it SExtractor} software.

\subsection{Astrometry and Photometric calibration}
For each CCD image, we used 10 standard stars from the USNO-B catalogue to find an astrometric solution. For all images, in both $R$ and $[$S$\scriptstyle\rm II$$]$ bands, we found the astrometric solution using the {\it ccmap} task of {\it IRAF}. In order to improve the accuracy of the astrometry, we used the {\it SCAMP} \citep{ber06} software which is a part of the TERAPIX pipeline. {\it SCAMP}  uses a larger number of star-like objects in the image to improve the astrometric accuracy. To stack all dithered images into a single final image, we used {\it SWARP} \citep{ber02}, also part of the TERAPIX software. In the co-addition process, {\it SWARP} uses all generated masks and the corresponding {\it SCAMP} astrometric solution. One of the advantages of {\it SWARP} is that it takes into account the variation of the signal to noise ratio of the images by allocating lower or higher weights during the co-addition process. Therefore, pixels with lower signal to noise ratios have correspondingly smaller weights. This preserves the Poissonian nature of the sky noise.

 We used two methods to find the zero point magnitude of the $R$-band images, namely the ``statistical method'', in which we used all star-like objects and the ``conventional method'', in which we only used the observed standard stars. Both methods are explained below.


i) In ``the statistical method", we first identified the stars in the $R$-band images using the {\it starfind} task in {\it IRAF}. To obtain the magnitude of the selected sources, we run {\it SExtractor} with a detection threshold of 2$\sigma$. Since SDSS covers the observed field, we also used SDSS DR7 to compare our resulting magnitudes in the $R$-band to those of SDSS. To transform the SDSS $ugriz$ magnitudes into Johnson $R$ magnitudes, we used the mean value of the following relations, which were presented by Lupton (2005)\footnote{For various transformations between SDSS $ugriz$ magnitudes to Johnson-Cousins $UBVRI$ system see \citet{jor06}, \citet{ive07} and \\{\tt \it http://www.sdss.org/dr6/algorithms/sdssUBVRITransform.html}}:

\begin{equation}R = r - 0.1837 (g - r) - 0.0971;  ~ \sigma = 0.0106\end{equation}
\begin{equation}R = r - 0.2936 (r - i) - 0.1439;  ~ \sigma = 0.0072\end{equation}

Estimating the median of the differences between the resulting {\it SExtractor} magnitudes and the corresponding SDSS magnitudes in $R$-band, for all star-like objects, helped us to estimate a more accurate zero point magnitude. As an example, Figure \ref{Rstack} shows the distribution of the magnitude shift, in one of the 10 WFC pointings of our survey. As seen, the $R$-band zero point magnitude is estimated to be $25.1\pm0.2$.

ii) In ``the conventional method'' we used the standard stars which were observed in all nights. In general, these stars were located in the center of the WFC field, free of vignetting. Four fields of standard stars from the catalogue of Landolt (1992) were taken for photometric calibration in this observation (i.e. SA98-1122, SA101-342, SA104-444 and SA107-612). We used {\it SExtractor} to obtain a catalogue of magnitudes and coordinates of the Landolt standard stars. We measured the zeropoint magnitude as $25.1\pm0.10$ for our $R$-band images which is in good agreement with the value obtained from the statistical method. To measure the flux of all detected objects in $R$-band, we run {\it SExtractor}.

\begin{table*}
\caption{Total integration times of the Final Stacked $R$ and $[$S$\scriptstyle\rm II$$]$
Images}
\label{lisint}
{\scriptsize
\begin{tabular}{ccccccccccc}
\tableline \tableline
Filter & F1 (SW) & F1 (SE) & F2 & F3 & F4 & 
CORE 1 & CORE 2 & CORE 3 & CORE 4 & CORE 5\\
  & (1)& (2)& (3)& (4)& (5)&
 (6)& (7)& (8)& (9)& (10)\\
\tableline
$R$ &  $9\times 300$  & $8\times 300$ & $8\times 300$ & $9\times 300$ & $5\times 300$ & $8\times 300$ &  $4\times 300$ & $8\times 300$ & $8\times 300$ & $8\times 300$ \\
 &  &  &  &  &  &  &   &   &   &   \\
$[$S$\scriptstyle\rm II$$]$& $9\times 1200$ & $9\times 1200$& $9\times 1200$ &  $9\times 1200$ & $6\times 1200$ & $10\times 1200$ & $9\times 1200$  & $9\times 1200$ & $9\times 1200$ & $11\times 1200$ \\
\tableline \tableline
\end{tabular}
}
\end{table*}

For photometric calibration of the narrow-band images in the $[$S$\scriptstyle\rm II$$]$ band, we convolved the spectrum of the spectrophotometric standard star (i.e. Hz21) with the throughput of the narrow-band ([S{\sc ii}]) and broad-band ($R$) filters. Comparing the expected magnitudes for the Hz21 star, in both filters, with those we have extracted from our imaging data, we were able to calibrate all the $[$S$\scriptstyle\rm II$$]$ band images.

To check the accuracy of our method for photometric calibration, we did two comparisons. We used $R$ magnitudes from the Adami et al. (2006) catalogue which had the most objects in common with our survey. Comparing Adami's $R$ magnitudes for all common objects with our corresponding $R$ magnitudes, we derived an RMS scatter of 0.19~mag and very good overall agreement, as can be seen in Figure \ref{fig:rcompare} (left panel). The next comparison was between the $R$ magnitudes of all detected galaxies in our observed field with those from the SDSS which is shown in Figure \ref{fig:rcompare} (right panel). The red line shows the completeness limit of the SDSS data; we find an RMS scatter of 0.22~mag up to this limiting magnitude. 

\section{ Measurements and results }
\label{Sec:Photometry}
\subsection{Scaling ratio and continuum subtraction}

\begin{figure*}
\begin{center}
\includegraphics[width=4in]{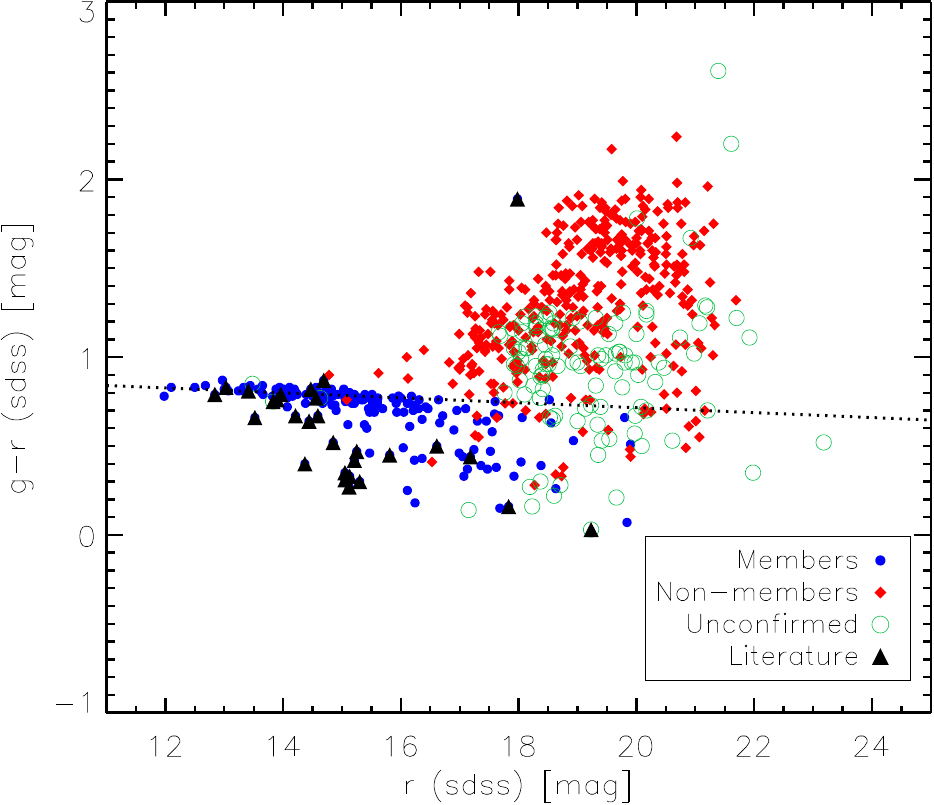}
\caption{Colour-magnitude diagram of all the galaxies detected in our continuum-subtracted images including both confirmed  (blue spots) and unconfirmed (green circles) cluster members as well as the H$\alpha$ emitting detected sources which are not Coma members based on their redshifts (red diamonds). Black triangles are 28 previously observed Coma members from the literature.}
\label{colmag}
\end{center}
\end{figure*}

An important step in measuring the H$\alpha$ flux is the removal of the continuum contribution to the flux in the images taken through the narrow-band $[$S$\scriptstyle\rm II$$]$ filter. The aim is to find the scaling factor to be applied to the $R$-band images to allow an accurate continuum subtraction.

We used a method which is based on the fact that the stars in the field should not present a net H$\alpha$ flux. Since we used the [S{\sc ii}] filter to detect the H$\alpha$ signal of the objects in the Coma cluster, all local foreground stars should be removed in the continuum subtraction process. In other words, an accurate scaling ratio should not leave any star visible in the continuum subtracted image. This allows us to obtain the mean scaling ratio over the surveyed field.

For small fields, a unique scaling ratio may result in reasonable measurements. Since we are dealing with a large field in this survey with varying signal-to-noise ratios, we used the scaling ratios which are allowed to vary across the field. To obtain accurate scaling ratios, first we found all stars in our observed field. Then, for any desired point in the observed field, we calculated the ratio of the fluxes in $R$ and $[$S$\scriptstyle\rm II$$]$-bands (i.e. $F_{R}/F_{\rm [S{\sc II}]}$) for all stars which are located within an aperture with a radius of 200 arcseconds centered at the point. In order to estimate the flux scaling ratio for each point, we calculated the median of $F_{R}/F_{\rm [S{\sc II}]}$ ratios of all stars within the aperture. In cases where insufficient stars were found within the 200$^{\prime\prime}$ aperture, we increased the size of the aperture to include at least 15 stars. In addition, we checked that the small changes in the size of the apertures and the minimum number of the stars would not significantly change the final estimated values. Figure \ref{ratiomap} shows the variation of the flux scaling ratio across the whole observed field. The H$\alpha$+[N{\sc ii}] flux and equivalent width (EW) of each object were calculated according to the estimated scaling ratio appropriate to its position in the field.

For each object, given the corresponding flux scaling ratio and its fluxes in both broad-band and narrow-band filters, we can measure its H$\alpha$ flux and Equivalent width (EW). We used the following expression to reach the H$\alpha$ (i.e. H$\alpha$ + [N{\sc ii}]) flux:

  \begin{equation} F_{H\alpha} = F_{[S{\sc II}]} - F_R /SR   \end{equation}

where SR is the associated scaling ratio which depends on the position of the object in the field (Figure \ref{ratiomap}), and $F_R$ and $F_{[S{\sc II}]}$ are the fluxes of the object in $R$ and $[$S$\scriptstyle\rm II$$]$ bands, respectively. After the continuum subtraction, the resulting image is called the H$\alpha$ image. The EW is calculated using narrow and broadband photometry. The observed EW is given by the following equation:

\begin{equation} 
EW= \frac{(F_{\rm[S{\sc II}] } \times SR - F_R) \times BW}{F_R  - F_{\rm [S{\sc II}]} } ; \\
BW = 80 \AA 
\end{equation}

where $BW$ is the band-width of the used narrow band filter (i.e. the [S{\sc II}] filter).

\begin{figure*}
\centering
\subfigure
{
\includegraphics[width=3in]{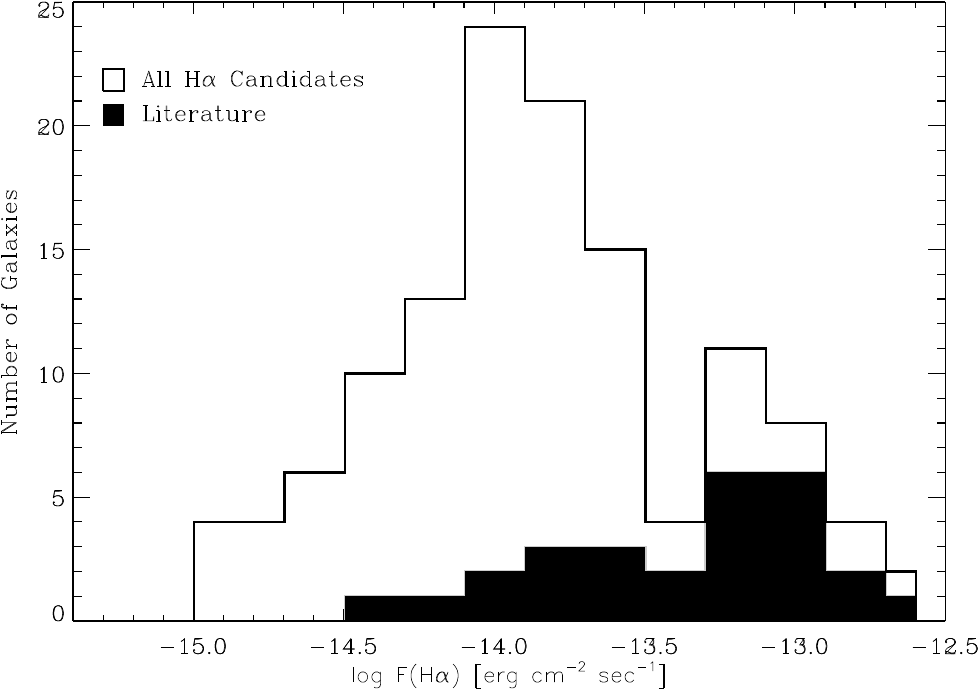}
}
\hspace{-0.1in}
\subfigure
{
\includegraphics[width=3in]{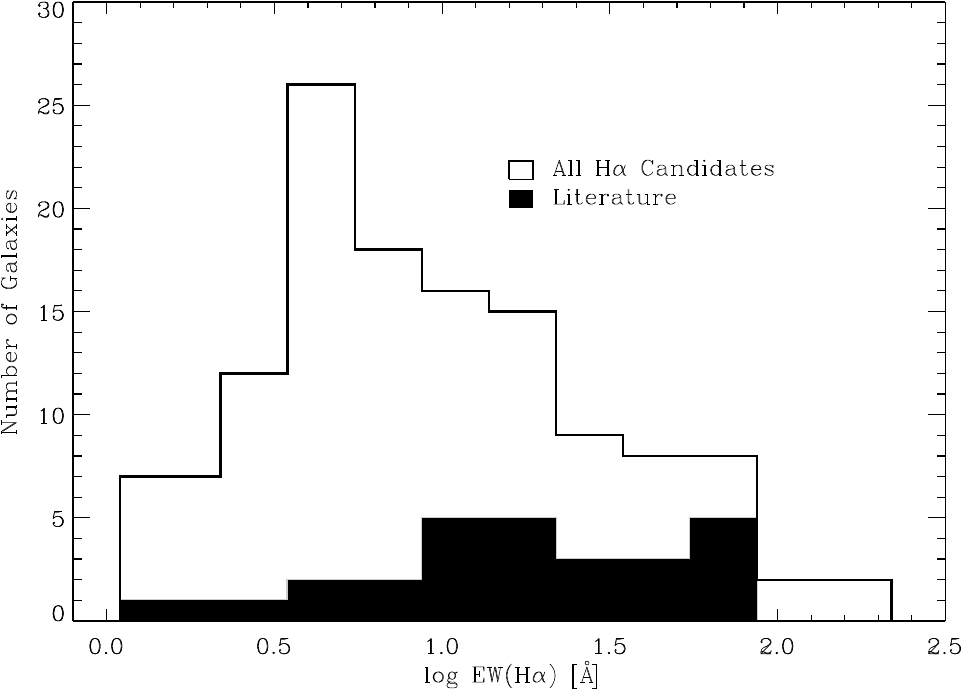}
}
\caption{The histogram of flux (left) and EWs (right) of the detected H$\alpha$ emitting sources in this survey. Those reported in previous observations are illustrated using the black solid regions. The open histograms are the fluxes and EWs of the sources detected in our survey. This study extends the faint detection limit by over one magnitude and as a result provides the largest sample ever reported.}
\label{histo}
\end{figure*}

H$\alpha$ emitting objects in our source catalogue were selected to have a signal-to-noise ratio larger than 1.5 in their H$\alpha$ fluxes as measured in the continuum-subtracted images. With this detection threshold, 124 H$\alpha$ emitting sources were found in total, with all previously observed and reported H$\alpha$ emitting sources in the Coma cluster detected in our images.

To check the membership of the detected H$\alpha$ emitting objects, we used different spectroscopic catalogues providing the Coma cluster membership, i.e., the complete spectroscopic sample of \citet{cal93}, the sample of low luminosity early type Coma galaxies of \citet{cal98}, the catalogue of radial velocities of \citet{col03}, catalogue of Coma cluster members of \citet{mob01} and \citet{kom02}, the SDSS sample of 196 Coma cluster members \citep{cas01}, SDSS Data Release 7 \citep{wes11}, and the Hectospec spectroscopy catalogue (Marzke et al. in preparation). Cross correlating our source catalogue with the membership catalogues, 124 H$\alpha$ emitting sources were found to be spectroscopically confirmed members of the Coma cluster, increasing the size of the sample of H$\alpha$ emitters of the Coma cluster by nearly a factor of five.

 For extended objects, the dominant source of errors is associated with the variations of the background on scales similar to the sources themselves. We have used {\it SExtractor} to measure fluxes within the Kron radii and the associated Poisson pixel counting errors. To derive the flux of the object, {\it SExtractor} subtracts the median of the background pixel values from all the pixels of the image. This ``background-subtracted'' image was produced for both broad-band and narrow-band images. 

The error of the flux is dependent on the Poisson noise arising from counting the number of pixels and the standard deviation of each individual pixel which is statistically constant over the final stacked image in each filter. As the numbers of frames stacked to produce the final images are not the same for all pointings, we used a weighting map to correctly add the contribution of each individual pixel to the photometry of each object in the final stack when estimating flux and EW errors. Ignoring this step would bias the estimated errors for objects extending over two pointings or in the overlapping regions. The measured properties of the objects in our final source catalogue are given in Tables \ref{litlis} and \ref{liscat}, listing respectively the properties of previously reported Coma cluster H$\alpha$-emitters, newly reported cluster H$\alpha$-emitters, and objects detected in our continuum-subtracted images but with unknown redshifts.

Figure \ref{colmag} shows the colour-magnitude diagram of all galaxies detected in our continuum-subtracted images which seem to have positive H$\alpha$ flux. Of these galaxies, 182 are confirmed (blue spots in Fig. \ref{colmag}) and 112 are unconfirmed (green circles in Fig. \ref{colmag}) cluster members. In addition, we have detected 358 H$\alpha$ emitting sources which are not Coma members according to their redshifts (red spots in Fig. \ref{colmag}). The number of 182 detected H$\alpha$ emitting sources reduces to 124 reported sources in the catalogue (Table 4) after applying the signal-to-noise ratio limit (i.e. 124 galaxies with F$_{H\alpha}$/Err(F$_{H\alpha}$)$>$1.5). Of these 124 galaxies, we have the H$\alpha$ flux for 28 galaxies from literature (Fig. \ref{colmag}, black triangles).

It is worth mentioning that non-cluster members detected in the continuum-subtracted images are not necessarily all emitting objects with emission lines coinciding with the $[$S$\scriptstyle\rm II$$]$ narrow-band filter. For a distant galaxy with no on-going star formation activity but with colours redder than those of Coma galaxies, the scaling ratio calibrated using local objects would artificially reduce their observed R-band luminosity relative to the light detected through the narrow-band filter, leading to a positive net flux in the continuum-subtracted images. The cleaning of the catalogue of emitting objects presented here from non-cluster members will be discussed in a forthcoming contribution.

\begin{figure*}
\centering

\subfigure
{
    \includegraphics[width=3.1in]{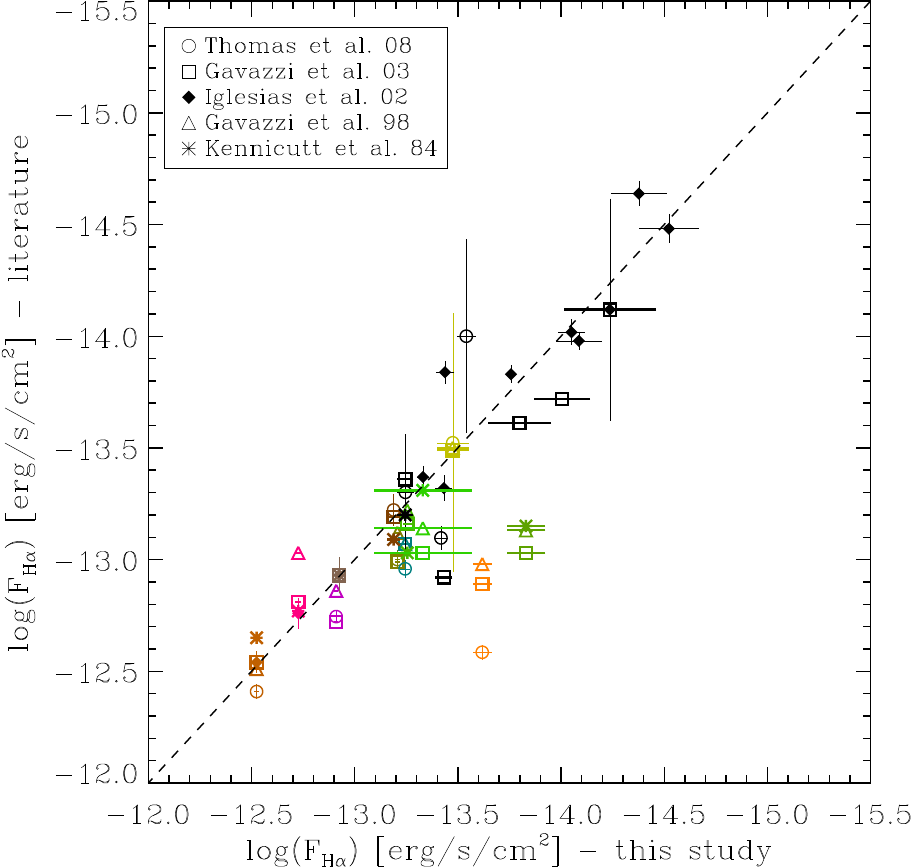}
}
\hspace{-0.1in}
\subfigure
{
    \includegraphics[width=3.1in]{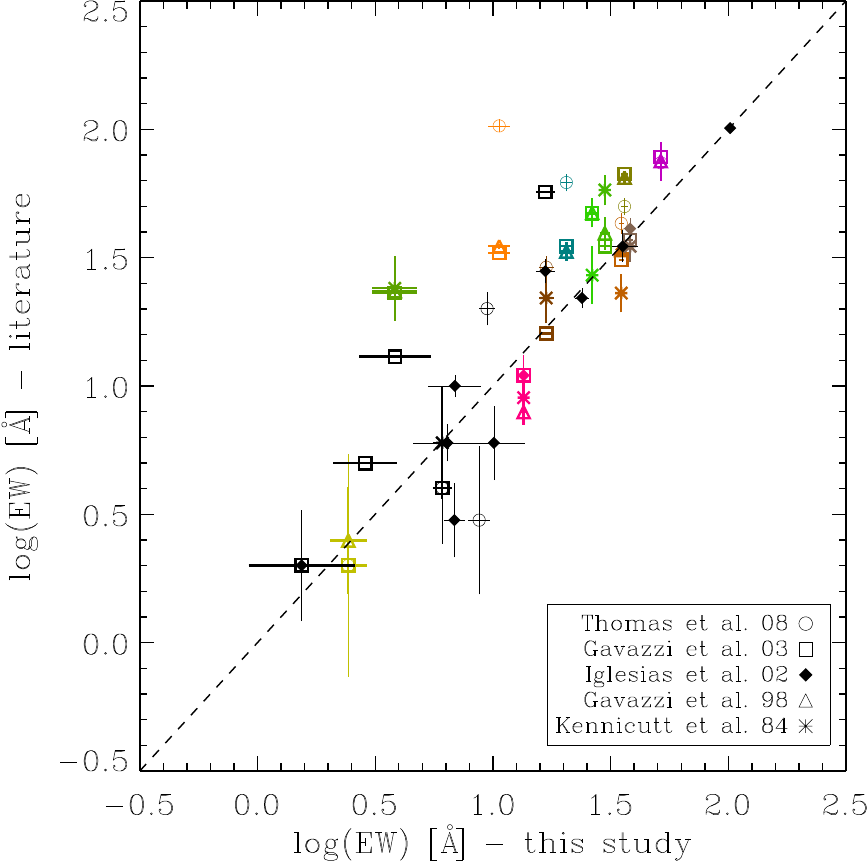}
}
\caption{Total H$\alpha$ fluxes and EWs obtained in this study versus the values in the literature for common objects. As seen, our results agree well with the measurements of \citet{igl02} which are carried out using same instrument and filter set. As there is more than one measurement for most of the galaxies in the sample, we have given a unique colour to each galaxy. Objects with only one reported value for the flux or the EW, are shown with black colour.}
\label{comp}
\end{figure*}

\subsection{Comparison with literature measurements, and deriving the H$\alpha$ catalogue}

We compared our results with existing catalogues of the H$\alpha$ observations in the Coma cluster. There are several objects in common with these studies; 9 objects with \citet{ken84}, 11 objects with \citet{gav98}, 12 objects with \citet{igl02}, 12 objects with \citet{gavv03}, 10 objects with \citet{tho08} and 7 objects with \citet{yag10}.

In Figure \ref{Rhisto} the histogram of $R$ magnitudes of Coma members reported in the literature (black solid regions) and of detected Coma members in this survey (open histogram) are illustrated. More faint members of the Coma cluster have been detected in our observation as it is deeper than previous surveys, except for that of \citet{yag10}, which covered a smaller observed area and focused on gaseous stripping events and extended emission line regions in the Coma cluster.

\begin{figure}

\hspace{-0.8cm}
\includegraphics[width=3.3in]{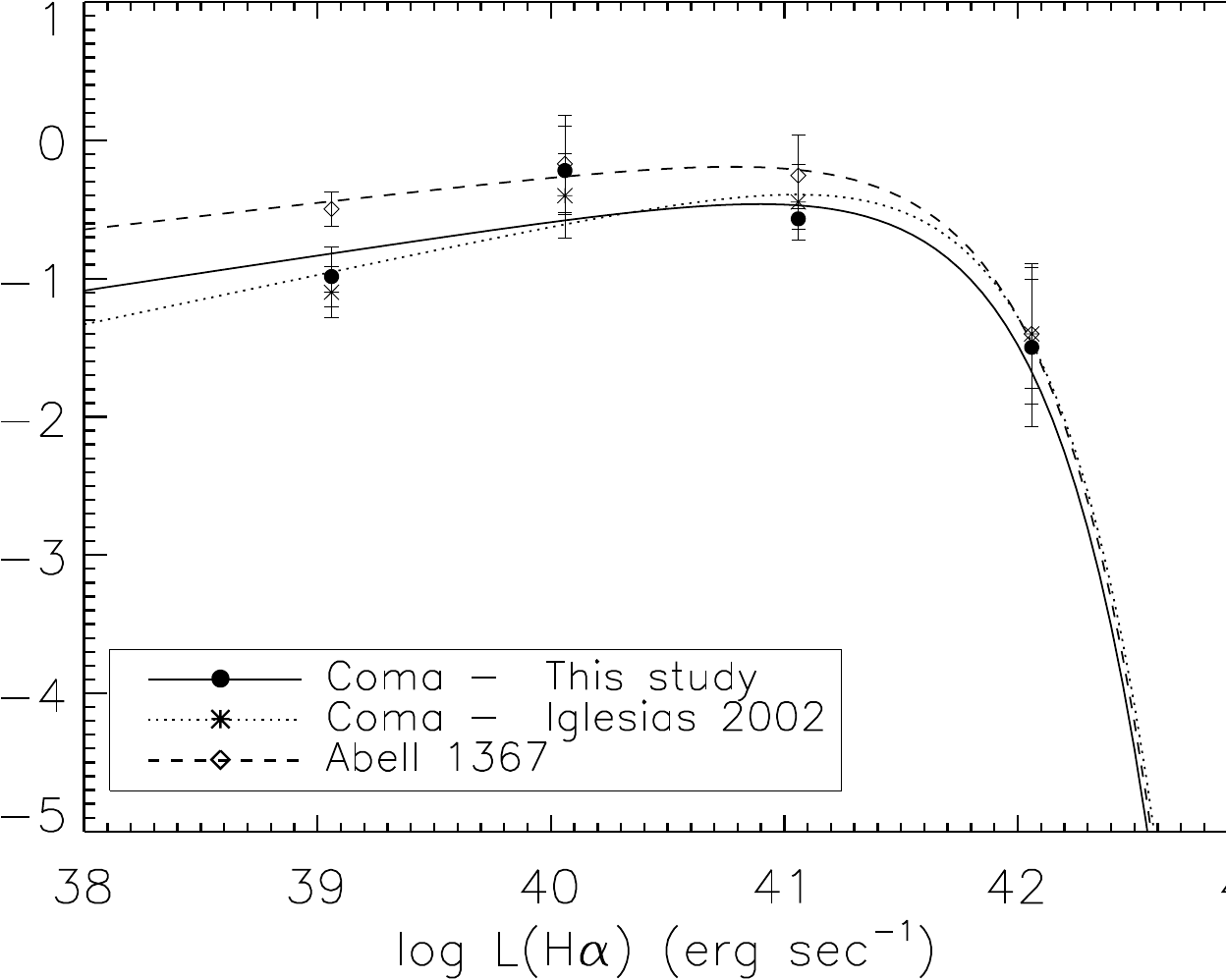}
\caption{The best fitted H$\alpha$ luminosity function for the Coma cluster based on the data in this study (filled cirrcles). The LFs for Coma (asterisks) and Abell 1367 (diamond) are from \citet{igl02}. All plots have been produced with the same fitting method. }
\label{LFfig}
\end{figure}

The distribution of the H$\alpha$ fluxes (left panel) and the EWs (right panel) of the detected H$\alpha$ emitting sources in this survey are illustrated in Figure \ref{histo}, where previously known sources are illustrated with the black solid regions and the open histogram represents all detected emitting sources in our survey. As seen in the left panel of the Figure \ref{histo}, our survey has extended the lower limit of the detected H$\alpha$-emitting objects to significantly fainter fluxes.
In Figure \ref{comp}, we compare our measured H$\alpha$ fluxes and EWs with the literature, for the common objects. Most of these objects in common have been observed more than once and therefore have more than one reported value for their fluxes and EWs (see Table \ref{litlis}).
Reported fluxes and EWs for each object have been shown with the same colour in both panels of the figure. Objects which have just one reported value in the literature, for flux or EW, have been shown with black symbols. The solid line shows a one-to-one correlation, which is a good fit to the data. There is a good overall correspondence between the different sets of measurements of fluxes and EWs. The mean zero points between our data and the literature agree very well. However, there is a sizable dispersion of 0.29 dex about the mean comparison of EW measurements. 

\begin{figure*}[h]
\begin{center}
\includegraphics[ width=0.8\textwidth,angle=0]{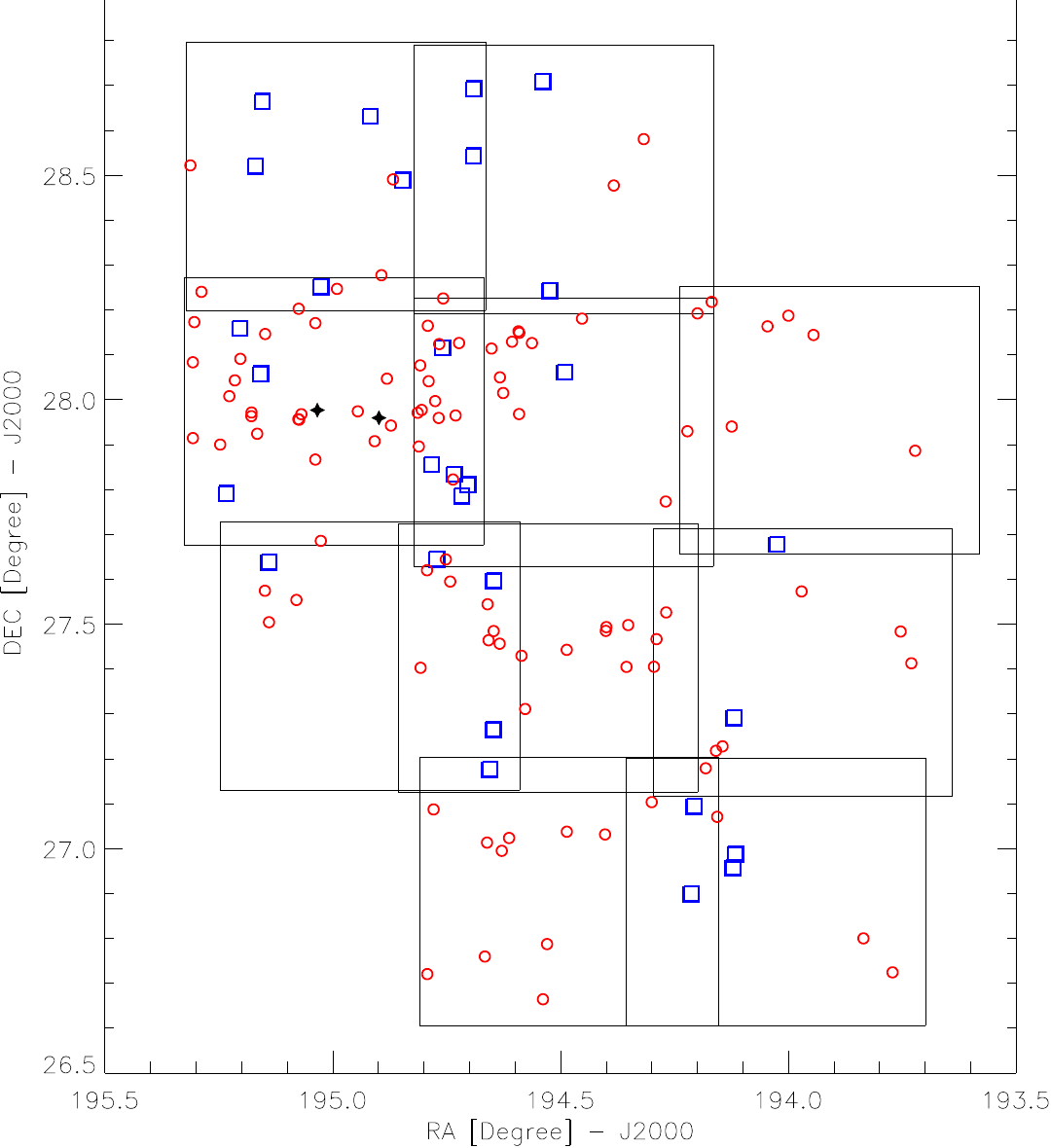}
\caption{The entire observed field and the WFC pointings are shown. The pointings from upper left to the right down are:  CORE 4, CORE 5, CORE 1, CORE 2, CORE 3, F1 (SE), F1 (SW), F2, F3 and F4, respectively (see Table \ref{lispo}). The final stacked H$\alpha$ image has similar area. Filled black symbols are NGC 4874 and NGC 4889, two ellipticals in the Coma cluster center. Blue squares are reported H$\alpha$ emitting sources in previous observations which we also detected and reported as H$\alpha$ emitters. Red circles are detected H$\alpha$ emitters in this survey which are confirmed by the spectroscopic surveys as Coma members.}
\label{poin}
\end{center}
\end{figure*}

Note that the scatter seen in both panels is artificially large due to the inclusion of different measurements for a number of objects in common with published catalogues. Both panels show that the best consistency between our measurements and those from the literature is with those presented in \citet{igl02}, who used the same observational setup as adopted in our survey. The agreement with the set of measurements from \citet{tho08}, is less satisfactory, however. Given the consistency of our measurements with those of \citet{igl02}, we could tentatively attribute this disagreement to systematic differences resulting from using different observational techniques.

The deviant objects with low fluxes and EWs that drive the dispersion seen in the figure, are all detected in our continuum-subtracted narrow-band images with low signal-to-noise ratios, i.e., S/N$\sim$2, rendering accurate measurements of narrow-band emission properties of those galaxies difficult. If nothing else this illustrates the challenging task in measuring accurate fluxes and EWs for faint H$\alpha$ emitting sources. If only objects detected above 10$\sigma$ in the continuum-subtracted narrow-band images are used in the comparison, the dispersion between our measurements and those in the literature decreases to 0.18 dex. Based on the comparisons against previous sets of measurements, we conclude that the methods we have used to subtract the continuum in the narrow-band images, to flux calibrate and measure our narrow-band imaging are not causing any systematic biases in our photometry.

In Figure \ref{poin}, the observed field and the WFC pointings are shown. The pointings (black squares) from upper left to the lower right are: CORE 4, CORE 5, CORE 1, CORE 2, CORE 3, F1, F2, F3, F4 respectively. The final stacked H$\alpha$ image has a similar area. The small blue squares are reported H$\alpha$ emitting sources in previous observations which are also detected in this survey. 

\begin{figure*}

\centering
\subfigure
{
    \includegraphics[width=2.0in]{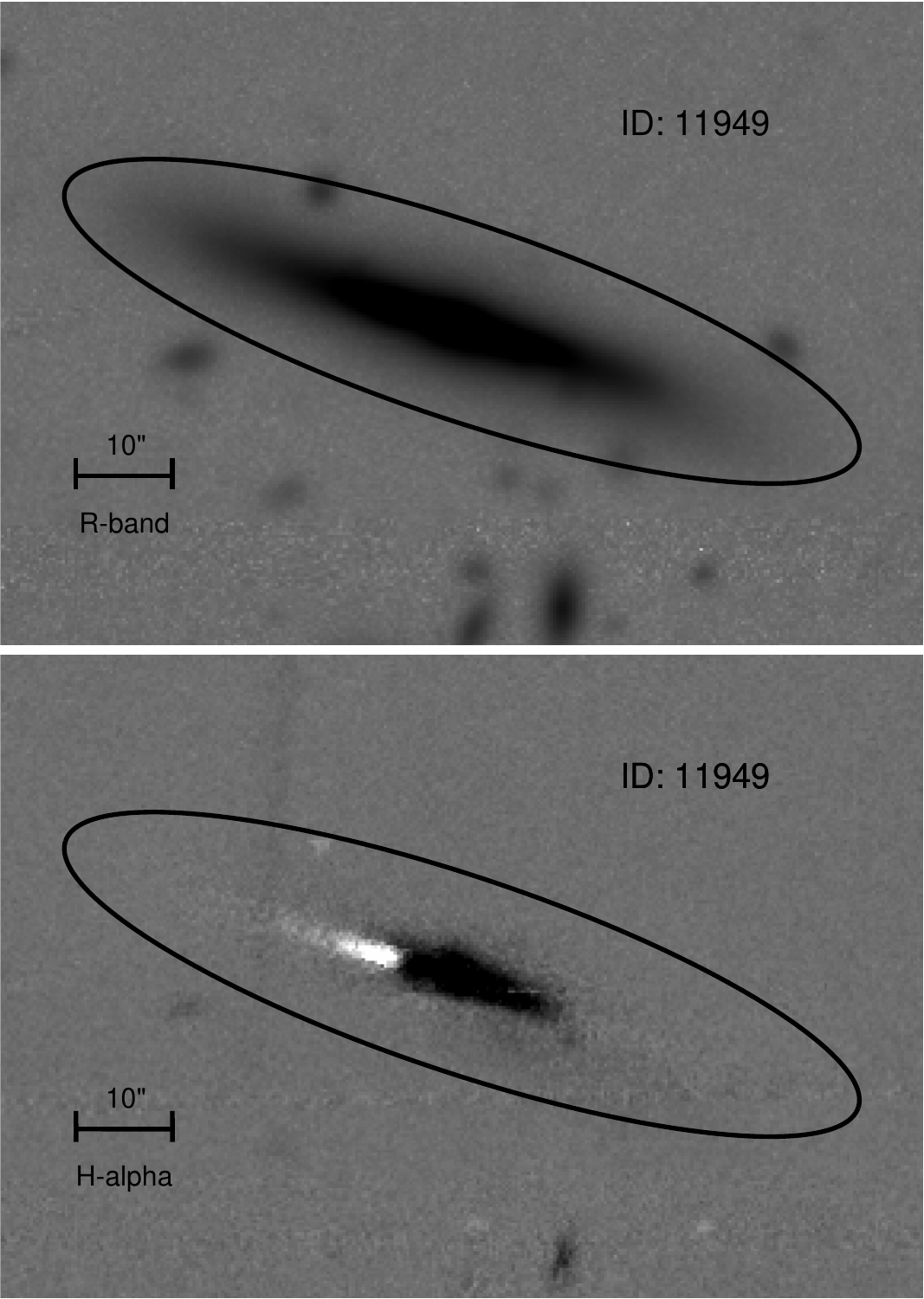}
}
\hspace{-0.1in}
\subfigure
{
    \includegraphics[width=2.0in]{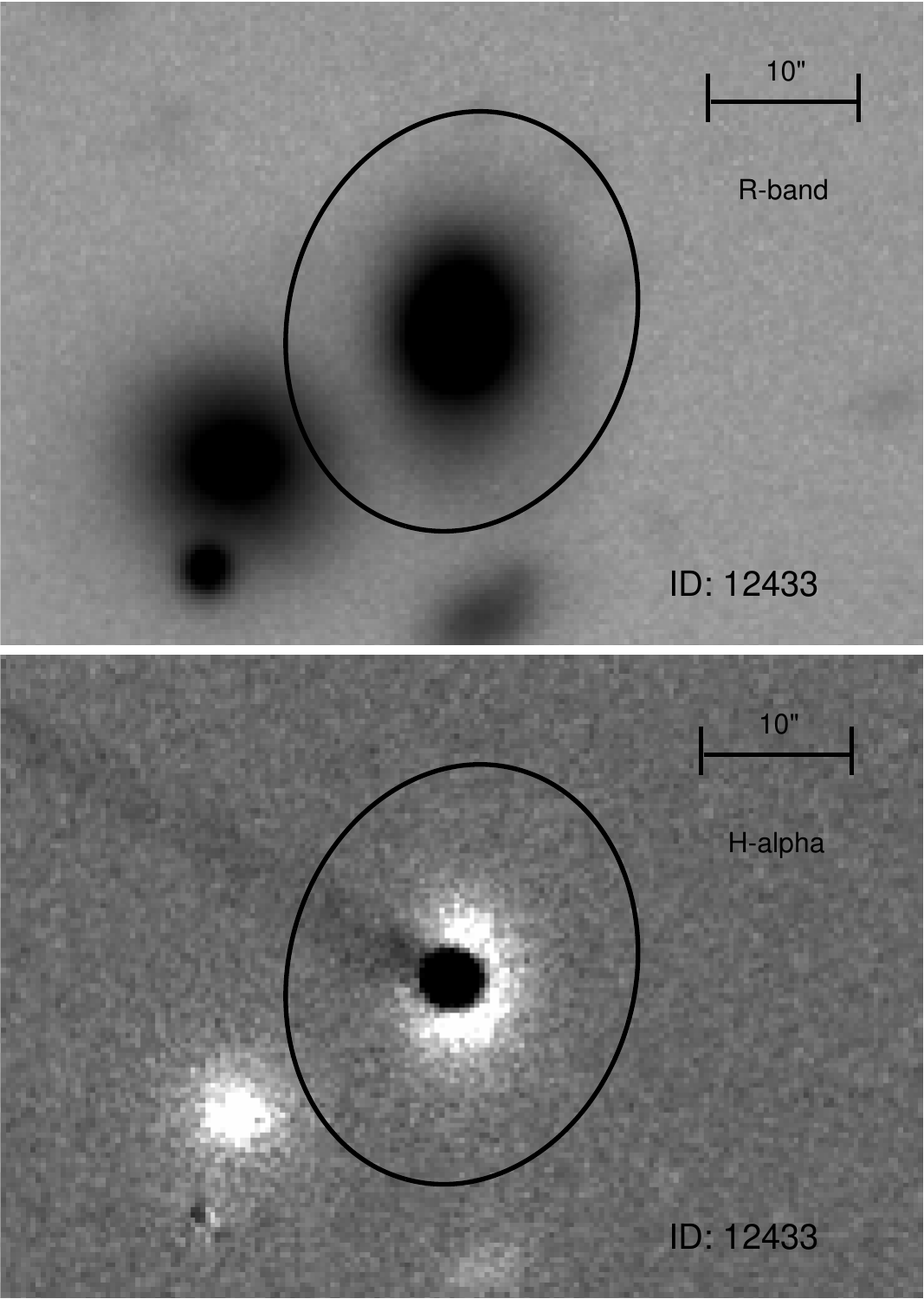}
}
\hspace{-0.1in}
\subfigure
{
    \includegraphics[width=2.0in]{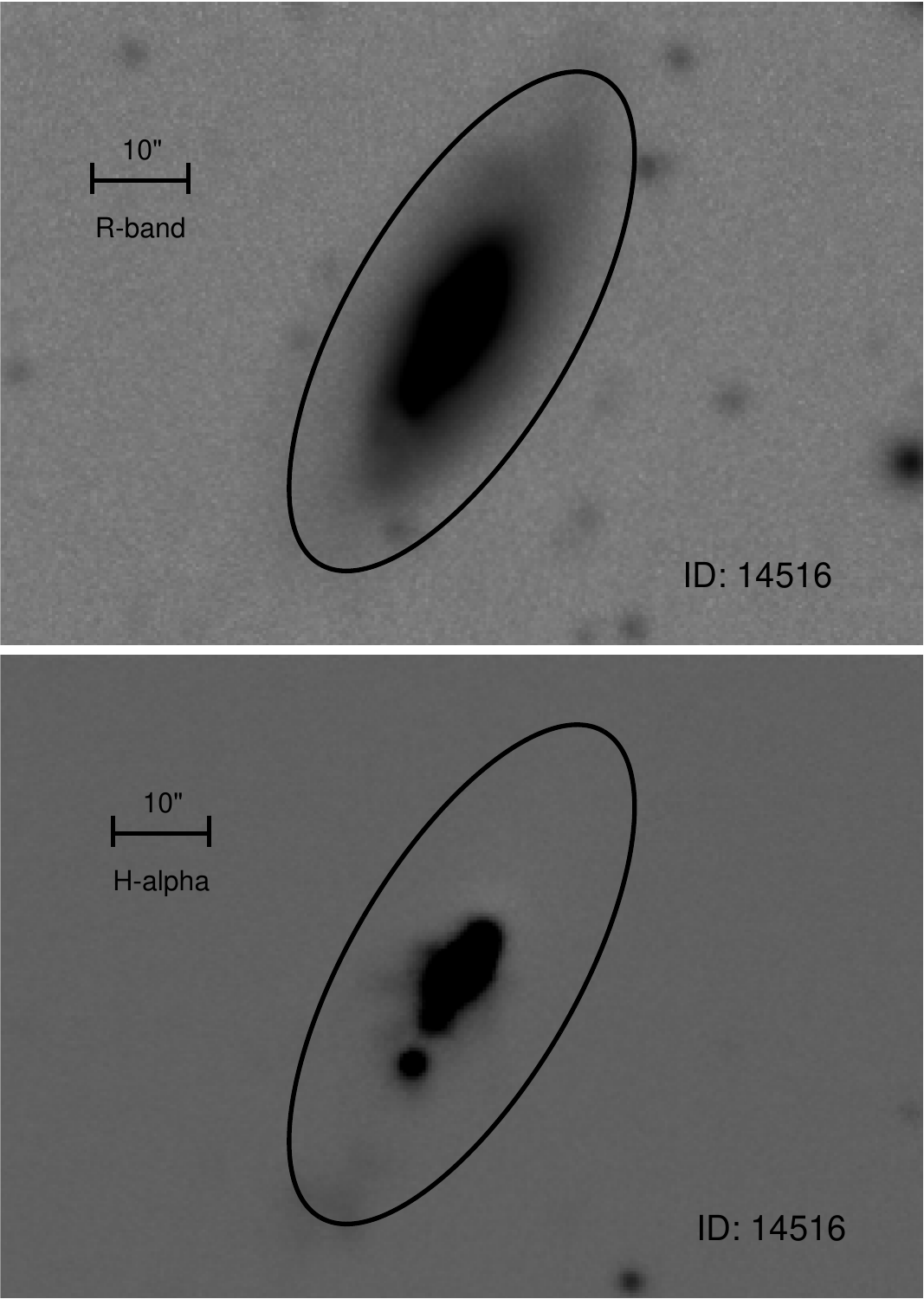}
}
\caption{Objects selected as H$\alpha$ emitting members of the Coma cluster which are previously detected. See Table \ref{litlis} for their measured and reported quantities in this study and in the literature. }
\label{secondpics}
\end{figure*}

\begin{figure*}

\centering
\subfigure
{
    \includegraphics[width=2.0in]{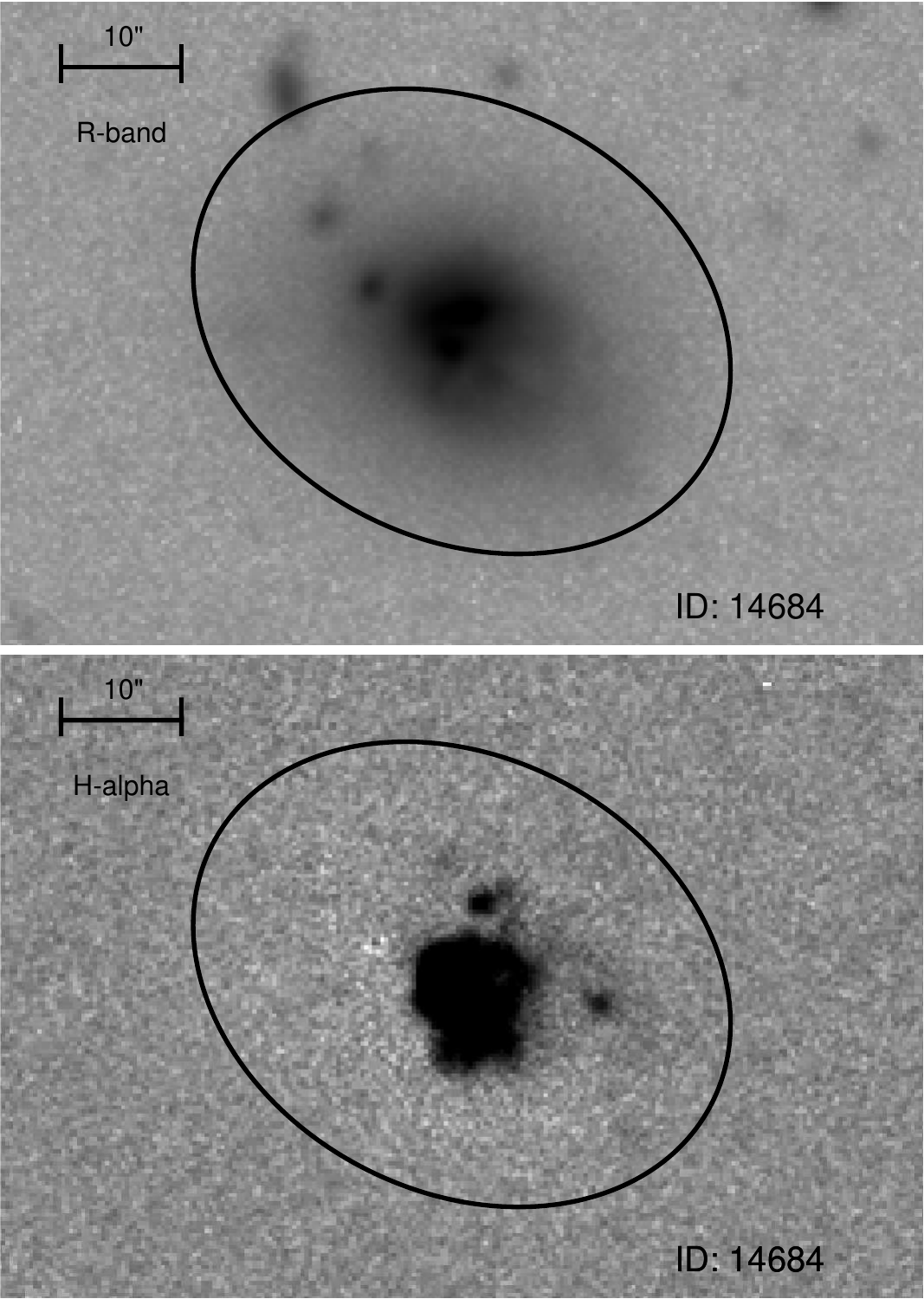}
}
\hspace{-0.1in}
\subfigure

{
    \includegraphics[width=2.0in]{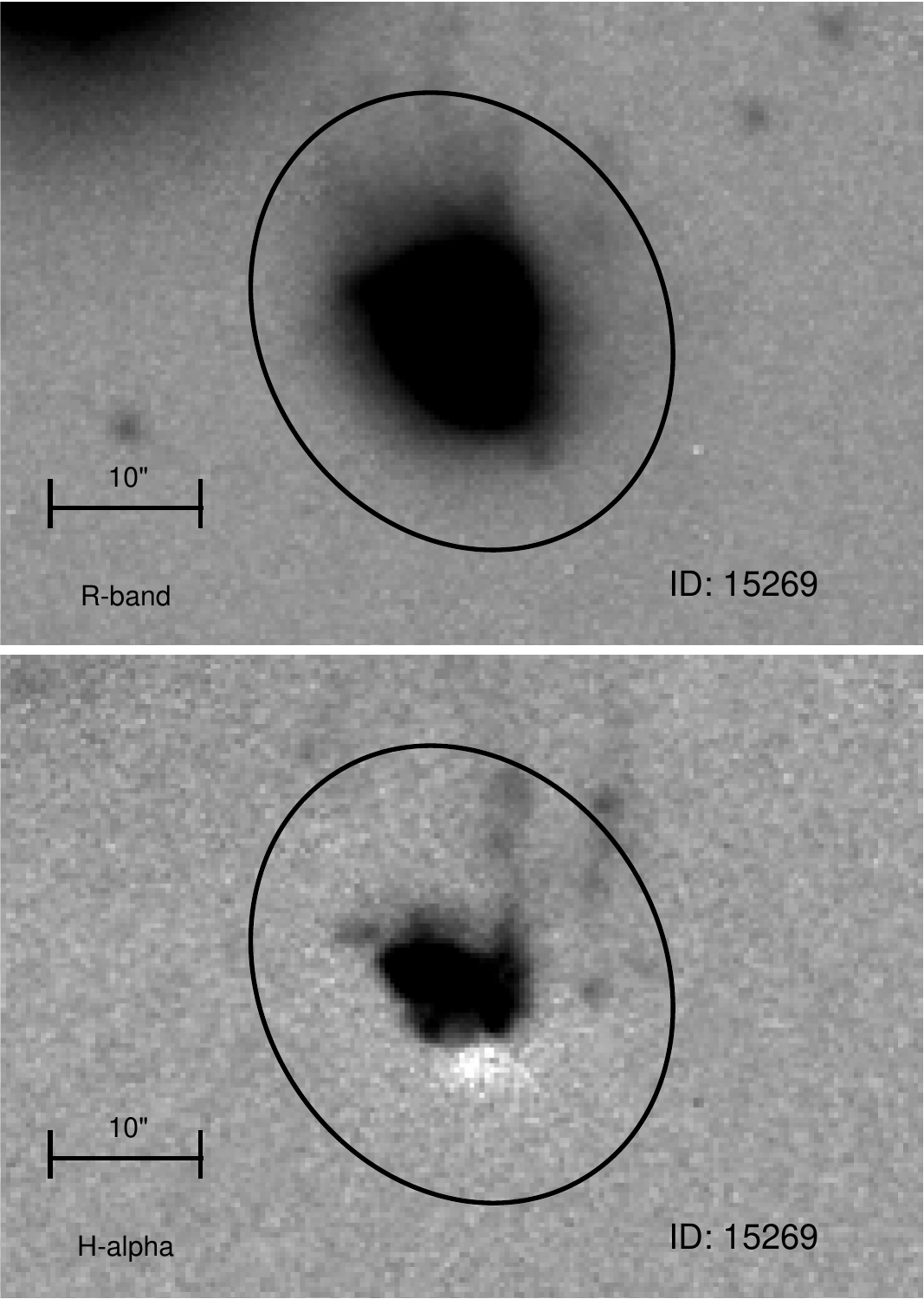}
}
\hspace{-0.1in}
\subfigure
{
     \includegraphics[width=2.0in]{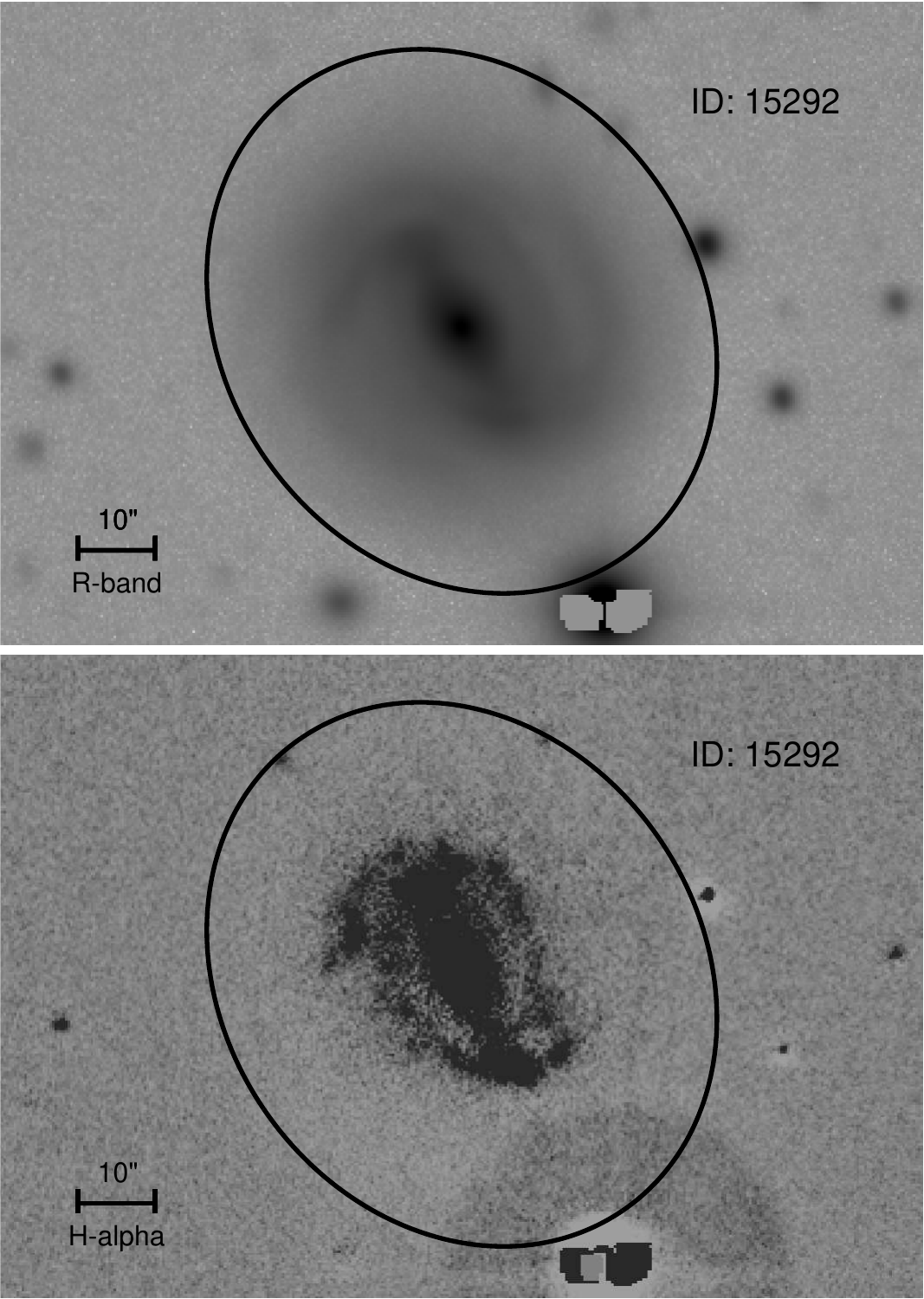}
}

\caption{Objects selected as H$\alpha$ emitting members of the Coma cluster which are previously detected. See Table \ref{litlis} for their measured and reported quantities in this study and in the literature.}
\label{thirdpics}
\end{figure*}

\begin{figure*}

\centering
\subfigure
{
    \includegraphics[width=2.0in]{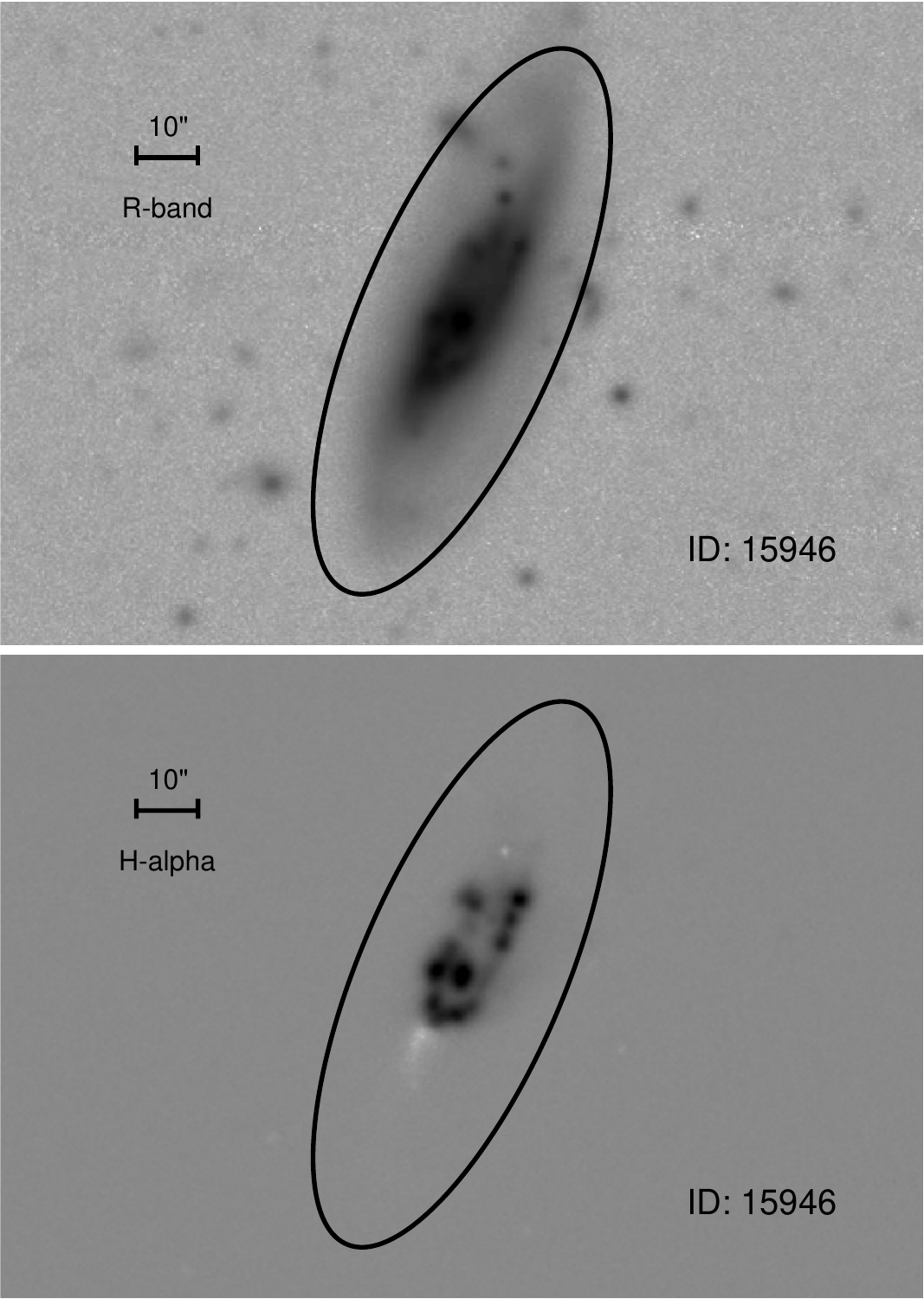}
}
\hspace{-0.1in}
\subfigure
{
    \includegraphics[width=2.0in]{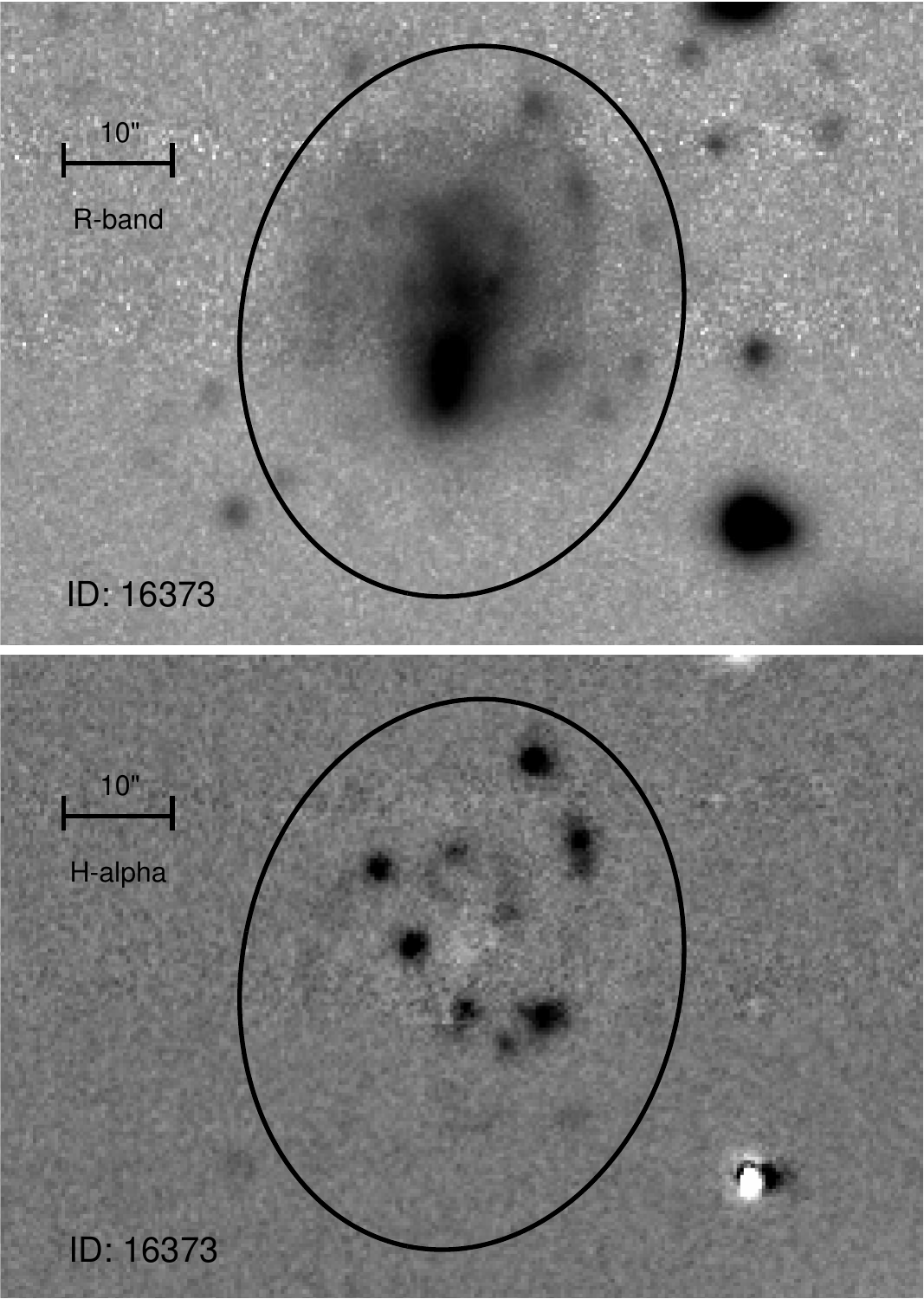}
}
\hspace{-0.1in}
\subfigure
{
    \includegraphics[width=2.0in]{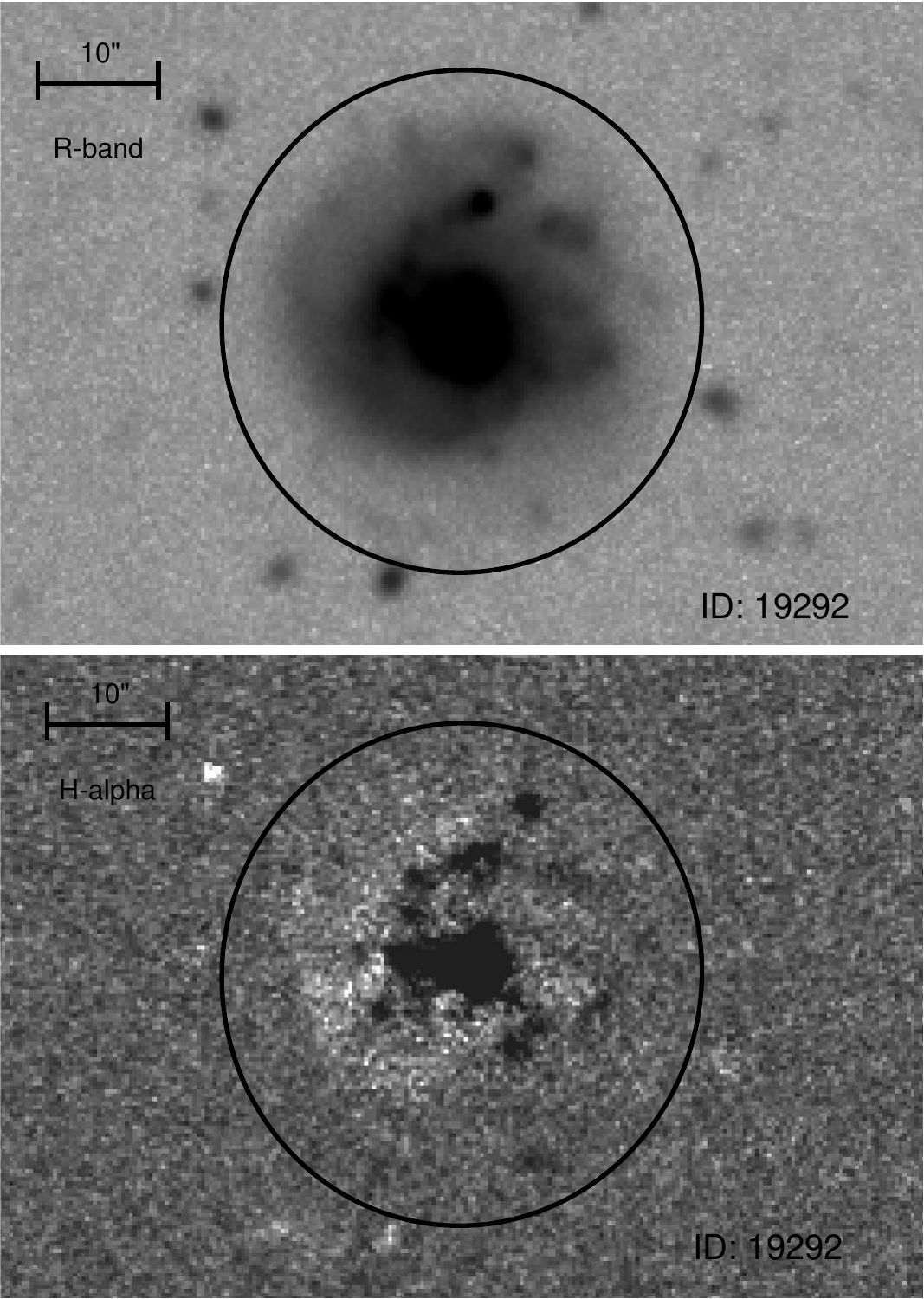}
}
\caption{Objects selected as H$\alpha$ emitting members of the Coma cluster which are previously detected. See Table \ref{litlis} for their measured and reported quantities in this study and in the literature.}
\label{forthpics}
\end{figure*}

 Figures \ref{secondpics}, \ref{thirdpics} and \ref{forthpics} show $R$-band (upper) and H$\alpha$ (lower) images of some previously observed Coma members which are also in the catalogue presented here. Our measured quantities and their reported values in the literature are listed in Table \ref{litlis}. 

Some galaxies show regions of apparent over-subtraction (negative values in the continuum-subtracted images). This could be due to colour variations that affect the continuum level, or to strong absorption features within the filter passband. These possibilities are being investigated through a programme of spectroscopic observations.

\subsection{The H$\alpha$ luminosity function}

Here, we present the H$\alpha$ luminosity function of all detected H$\alpha$ emitting members of Coma cluster in this sample. The measured H$\alpha$ fluxes were corrected to remove the effects of contamination from [NII] emission and internal extinction. The [NII] contamination correction was taken from \citet{hel04}, which introduce a relation between R-band luminosity and [NII]/H$\alpha$ in the following form

\begin{equation}
log\frac{[NII]}{H\alpha}=(-0.13\pm0.035)M_R+(-3.2\pm0.90),
\end{equation}

where $M_R$ is the absolute magnitude of H$\alpha$ emitting members in the R-band. \citet{hel04} also evaluated a type-independent dust extinction correction relation as

\begin{equation}
logA(H\alpha)_{int}=(-0.12\pm0.048)M_R+(-2.5\pm0.96).
\end{equation}
 
In order to make a proper comparison of this study with the previous H$\alpha$ luminosity function study of the Coma cluster \citep{igl02}, we also assume an angular radius of 4 degrees for Coma cluster. 

Here, we adopted a single Schechter function \citep{sch76} for the H$\alpha$ luminosity distribution of Coma members in the following form

\begin{equation}
\phi(L)dL=\phi^*(L/L^*)^{\alpha}exp(-L/L^*)d(-L/L^*).
\end{equation}

where $\phi^*$ is the normalization parameter, $L^*$ is the characteristic luminosity and $\alpha$ refers to the faint-end slope of the Schechter Function. Considering the uncertainty arising due to small number of detected members, we chose the binning interval in the range 38.5 to 42.5 for Log L(H$\alpha$) with the bin size $\delta$logL=1.0. This selection minimizes the Poissonian uncertainties of galaxy count in each bin. In addition, we used the same fitting method to evaluate the H$\alpha$ LF Schechter parameters for Coma and Abell 1367 based on the data extracted from \citet{igl02}. 
\begin{table}[h]
\renewcommand{\arraystretch}{0.99}\renewcommand{\tabcolsep}{0.15cm}
\caption{Best fitting Schechter parameters and related errors for H$\alpha$ emitting sources in Coma and Abell 1367, presented in this study and in \citet{igl02}.}
\label{LFtable}
\begin{center}
\begin{tabular}{l c c }
\tableline 
 Sample  &   $\alpha$  &   log(L$^*$) \\
   &    &  erg s$^{-1}$ \\

\tableline  \\
Coma (this study) & -0.75$\pm$0.13 &  41.49$\pm$0.24  \\ 
Coma (Iglesias 2002) & -0.64$\pm$0.17 &  41.50$\pm$0.21  \\ 
Abell 1367 (Iglesias 2002) & -0.81$\pm$0.11 &  41.49$\pm$0.17  \\ 
\tableline 
\end{tabular}
\end{center}
\end{table}
 
Table \ref{LFtable} shows the best fitting Schechter parameters and related errors for H$\alpha$ emitting members, presented in this study and in \citet{igl02}. In our calculations, we assume that H$_0$=71 km s$^{-1}$ Mpc$^{-1}$. To compare our resulting H$\alpha$ LF with those presened in \citet{igl02}, we derived the fitting Schechter parameters for \citet{igl02} galaxies with H$_0$=71 km s$^{-1}$ Mpc$^{-1}$. Figure \ref{LFfig} shows the Schechter LFs of all samples and reveals a full agreement between these studies. 
 
We confirm the shallow faint-end slope, $\sim$-0.7, found for Coma by \citet{igl02}. A similar slope was found by \citet{ume04} for the H$\alpha$ LF of the z=0.25 cluster Abell 521. These slopes imply far fewer star-forming dwarf galaxies than is found in field populations, with typical faint-end slopes in field H$\alpha$ LFs being in the range -1.3 to -1.7 \citep{gal95, tre98, sul2000, fuj2003, hip2003, ly2007, wes2008}.  \citet{jam2008} show that a steep faint-end slope of $\sim$-1.4 for the Ha LF of field galaxies extends far into the faint dwarf regime, implying a large population of star-forming dwarfs that is suppressed in the high-density environment of the Coma cluster. The intermediate slope of -0.85 for the Ha LF in Abell 1367 \citep{igl02} indicates that the suppression of star formation in the dwarf galaxies has been less effective in this cluster, which is known to be less dynamically-evolved than Coma.

\section{Summary }
\label{Summary}
A deep and wide-field narrow-band imaging survey of the local Coma cluster, the prototype of local rich and relaxed galaxy clusters, obtained with the WFC/INT is presented. The survey is aimed at measuring accurate narrow-band photometry of the Coma cluster galaxies down to the dwarf regime over a region extended from the cluster core to the infall region. Our survey is sampling galaxies covering extended ranges of galaxy morphology, luminosity, and environment.
 124 out of the 185 detected objects in our deep continuum-subtracted images have spectroscopically determined cluster membership, 80\% of which have measured H$\alpha$ emission line properties for the first time. Our estimated H$\alpha$ luminosity function of the Coma cluster agrees well with what previously obtained by \citet{igl02}. The analysis of the properties of the ongoing (unobscured) star formation activity of the cluster galaxies and how they change as a function of their properties, using the set of observations presented here combined with our ongoing multi-wavelength survey of the Coma cluster, will be presented in a series of forthcoming papers.

\acknowledgements

\section{Acknowledgments}
{\it SE} would like to thank Daniel J. Smith for providing confidence map for data reduction. The authors wish to acknowledge The Isaac Newton Telescope operating Group in the Spanish Observatorio del Roque de los Muchachos of the Instituto de Astrof\' isica de Canarias.

\clearpage

\clearpage
\onecolumn

\begin{table}

\renewcommand{\arraystretch}{0.99}\renewcommand{\tabcolsep}{0.1cm}

\caption{Objects selected as H$\alpha$ emitting members of the Coma cluster which also have been detected in previous observations.
}

\footnotesize
\centering
\label{litlis}
\begin{tabular}{c c c c| c c| c c |c c c c}

\tableline
ID  & Other names & RA & Dec. & H$\alpha$ Flux & H$\alpha$ Flux (lit.{*}) & EW & EW (lit.)& $R$ magnitude & Type & z & Ref.{**}\\
 &  & J2000 & J2000 & \multicolumn{2}{c|}{(10$^{-14}$  erg cm$^{-2}$s$^{-1}$)} & ($\AA$) & (mag) & & & \\ 
\tableline
\tableline
1750 & 160-033 & 194.2132 & 26.8989 & 3.82$\pm$0.30& 8.0$\pm$1.0 & 9.44$\pm$0.73 &20.0$\pm$3.0 & 14.2&SA0/a &0.021 &5\\
&  & & & & & & & &  & &\\

2105 & CGCG160-025 & 194.1161 & 26.9874 & 3.35$\pm$0.60 & 3.16 & 2.43$\pm$0.44& 2.50$\pm$1.2& 12.87 & SBa/b & 0.022&2\\
  &  CGCG160025  & &  & &  3.24   & &  2     & & & & 4 \\
& 160-025& & & &3.0$\pm$4.0  & & 2.0$\pm$2.0& & & &5\\
&  & &  & & & & & & & &\\

2209 & CGCG 160-024 & 194.1214 & 26.9570 & 0.53$\pm$0.39 & 0.21 & 0.87$\pm$0.64& 0.4$\pm$3.2&13.75 &SA0 &0.024 &2\\
& & &  & & & &  & & & &\\

3195 & 160-031 & 194.2069 & 27.0939 & 2.88$\pm$0.31 & 1.0$\pm$1.0 &8.74$\pm$0.92 &3.0$\pm$2.0 &14.42 &SA0/a &0.023 &5\\
& & &  & & & & & & & &\\

3913&CGCG 160-067 & 194.6553 & 27.1766 & 12.30$\pm$0.24 & 13.8 &51.64$\pm$1.03 &75.1$\pm$13.20 &14.77 & Sc & 0.026 &2\\
  &  CGCG160067   & & & &  19.05   & &  78     & & & & 4 \\
&160-067 & & & & 18.0$\pm$1.0 & &74.0$\pm$5.0 & & & &5\\
& & &  & & & & &   & & &\\

5001 & CGCG 160-064 &194.6473 & 27.2647 & 6.21$\pm$0.20 & 7.59 &36.22$\pm$1.15 & 64.80$\pm$3.7 & 15.11 &S & 0.025&2\\
  &  CGCG160064 & &   & &  10.23   & &  67     & & & & 4 \\
& 160-064 & & & & 10.0$\pm$1.0 & &50.0$\pm$4.0 & & & &5\\
& & &  & & & & &  & & &\\

5174 & CGCG 160-026 & 194.1192 & 27.2914 & 5.70$\pm$0.31& 8.51 & 20.51$\pm$1.11& 33.4$\pm$2.90&14.6 &SA &0.025 &2\\
  &  CGCG160026  & &  & &  8.51   & &  35     & & & & 4 \\
&160-026 & & &  & 11.0$\pm$1.0 & &62.0$\pm$5.0 & & & &5\\
& & &  & & & & &  & & &\\

9040 & 160-068 & 194.6466 & 27.5964 & 5.68$\pm$0.52 & 5.0$\pm$3.0 & 6.09$\pm$0.56 & 4.0$\pm$2.0 & 13.3 & Ep& 0.026 & 5\\
 & GMP 4156 & & & & 0.065 & & - & & & &6\\
& & &  & & & & & &  & &\\

9508 & Z160-073 & 194.7719 & 27.6443 & 1.48$\pm$0.32 & 7.08 &3.83$\pm$0.84 & 24.0$\pm$7.0& 14.25 & SBb &0.018 & 1 \\
 & CGCG 160-073 & & &   & 7.4 & &23.4$\pm$1.37 & & & &2\\
  &  CGCG160073    & & &  &  9.33   & &  23     & & & & 4 \\
 & GMP 3779 & & &   & 3.7& & - & & & & 6\\
& &  &  & & & & & &  & &\\

9625 & Z160-086 & 195.1404 & 27.6377 & 5.56$\pm$0.28 & 9.33 &29.87$\pm$1.5 &58.0$\pm$8.0 & 15.06& Sc& 0.025&1\\
& CGCG 160-086 & & &  & 6.03 & &39.3$\pm$5.8 & & Irr& &2\\
  &  CGCG160086  & &  & &  6.92   & &  35     & & & & 4 \\
& & &  & & & & & & & &\\

9965 & CGCG 160-020 & 194.0254 & 27.6779 & 2.41$\pm$0.26 & 10.47 & 10.63$\pm$1.13&35.0$\pm$0.8 &14.86 &SBa & 0.016&2\\
  &  CGCG160020   & & & &  12.88   & &  33     & & & & 4 \\
& 160-020 & & & & 26.0$\pm$2.0 & &103.0$\pm$6.0 & & Sa& &5\\
& & &  & & & & &  & & &\\

10979 & NGC 4911 & 195.2335 & 27.7909 & 18.78$\pm$0.67 &17.0 &13.5$\pm$0.48 & 9.0$\pm$2.0& 12.86 & SAb&0.027 &1\\
 & CGCG160-260 & & &  & 9.33&  &7.9$\pm$0.9 & & & & 2\\
& 130056+274727 & & & & 17.38$\pm$2.80 &  & 11$\pm$2.0 &  & & & 3\\
  &  CGCG160260   & & & &  15.49   & &  11     & & & & 4 \\
& GMP2374 & & & & 23.0 & & - & & & &6\\
&  &  & & & & & &  & & &\\

11949  & IC3949 & 194.7332 & 27.8333 & 3.65$\pm$0.37 & - & 6.86$\pm$0.69 &0$\pm$2.0 &13.87 &Sb & 0.025 &1 \\
 & 125856+275002 & & & & 1.45$\pm$0.17& & 3.0$\pm$1.0 & & & &3\\
 & GMP 3896 & & & & 3.9 &  & - &  & & & 6\\
&  &  & & & & & &  & & &\\

12208 & 125907+275118 & 194.7831 & 27.8550 & 0.58$\pm$0.30 & 0.76$\pm$0.87 &1.54$\pm$0.8 &2.0$\pm$1.0 & 14.28& SB0/a & 0.022 & 3\\
  &  CGCG160219   & &    & &  0.76   & &  2     & & & & 4 \\
& & &  & & & & &   & & &\\

12433  &  CGCG160243 & 195.0380 & 27.8664 & 1.59$\pm$0.56 &  2.45   & 3.84$\pm$1.34&  13     & 15.17 & Irr/S & 0.018& 4 \\
  &  GMP 2910   & &    & &  1.9   & &  -     & & & & 6 \\
 & &  & & & &  & &  & & &\\

12672  &  CGCG160261 & 195.2469 & 27.8998 & 0.99$\pm$0.31 &  1.91   &2.87$\pm$0.9 &  5     &14.35 & Irr/S &0.023 & 4 \\
 & &  & & &  & &   & & & &\\
\tableline
\multicolumn{10}{r}{{Continued on next page}} \\

\end{tabular}
\end{table}

\setcounter{table}{3}
\begin{table}

\renewcommand{\arraystretch}{0.99}\renewcommand{\tabcolsep}{0.1cm}

\caption{Continued from previous page
}

\footnotesize

\label{litlis}
\begin{center}
\begin{tabular}{c c c c| c c| c c |c c c c}

\tableline
ID  & Other names & RA & Dec. & H$\alpha$ Flux & H$\alpha$ Flux (lit.{*}) & EW & EW (lit.)& $R$ magnitude & Type & z & Ref.{**}\\
 &  & J2000 & J2000 & \multicolumn{2}{c|}{(10$^{-14}$  erg cm$^{-2}$s$^{-1}$)} & ($\AA$) & (mag) & & & \\ 
\tableline
\tableline

14516 & IC 4040 & 195.1579 & 28.0575 & 11.90$\pm$0.35 & 11.75 & 38.25$\pm$1.13 &35.0$\pm$5.0 &14.49 & SAb & 0.026 & 1\\
 & 130037+280327 &  & &   & 11.75$\pm$2.16 & & 41.0$\pm$4.0 & & & & 3\\
  &  CGCG160252   & &    & &  11.75   & &  37     & & & & 4 \\
 & GMP 2559 &  & &   & 11.0 &  & - &  & & & 6\\
 & &  & & & & & &   & & &\\

14684 & 125757+280343 & 194.4907 & 28.0615 & 4.67$\pm$0.33 & 4.27$\pm$0.49 & 23.9$\pm$1.71 & 22.0$\pm$2.0 & 14.98 & Irr/S & 0.027 & 3\\
 & &  & & & & & &   & & &\\

15269 & 125902+280656 & 194.7586 & 28.1156 & 3.70$\pm$0.35&4.79$\pm$0.64 & 16.7$\pm$1.56 & 28.0$\pm$3.0& 14.88 & SBb & 0.031&3 \\
  &  CGCG160213  & &     & &  12.02   & &  57     & & & & 4 \\
 & GMP 3816 &  & &   &10 & & - & & & &6\\
 & &  & & & &  & &  & & &\\

15292 & NGC 4907 & 195.2034 & 28.1584 & 5.69$\pm$0.52 & 6.31 & 6.09$\pm$0.56 &6.0$\pm$3.0 & 13.33 & Sb & 0.019 &1\\
  &  CGCG160257  & &     & &  4.37   & &  4     & & & & 4 \\
& &  & & & & &  & &   & &\\

15946 & NGC 4848 & 194.5234 & 28.2425 & 29.9$\pm$0.79 & 22.39 &35.09$\pm$0.93 &23.0$\pm$4.0 &13.34 & Sc & 0.024 &1\\
 & CGCG 160-055 &      & &             & 30.9   &       & 33.8 $\pm$1.3&  &  && 2\\
 & 125805+281433 &  & &   &28.84$\pm$3.32 & &34.0$\pm$2.0 & &  & & 3\\
  &  CGCG160055    & &   & &  28.84   & &  31     & & & & 4 \\
& 160-055 &  & &   & 39.0$\pm$3.0 &  & 43.0$\pm$4.0&  &  & &5\\
& &  & & & & & &   & & &\\

16373 & 130006+281500 & 195.0258 & 28.2514 & 0.30$\pm$0.10 & 0.33$\pm$0.05 & 6.41$\pm$2.16 & 6.0$\pm$1.0 & 16.53& SBc& 0.021 & 3\\
 & &  & & & & &   & & & &\\

18249& 125923+282919 & 194.8464 & 28.4886 & 0.82$\pm$0.21 & 1.05$\pm$0.10 &6.89$\pm$1.79 &10.0$\pm$1.0 &15.54 &  SA0/a & 0.023&3\\
& & &  & &   & & & & & &\\

18782& 125845+283235 & 194.6910 & 28.5430 & 1.75$\pm$0.09& 1.48$\pm$0.14 &101.77$\pm$5.28 &101.0$\pm$4.0 &17.55 & SA & - &3\\
& & &  & &   & & & & & &\\

19292 & Z160-076 & 194.9174 & 28.6309 & 4.68$\pm$2.56 &4.90 & 26.4$\pm$1.5 & 27.0$\pm$7.0 & 15.14 &Sc& - & 1 \\
 & CGCG 160-076  & &  &  &7.24 & & 47.20$\pm$6.1 & & & &2\\
  &  CGCG160076   & &    & &  9.33   & &  47     & & & & 4 \\
 & &  & & &  & &   & & & &\\

19541&130037+283951 & 195.1548 & 28.6642 & 0.42$\pm$0.13 & 0.23$\pm$0.03 & 10.1$\pm$3.11& 6$\pm$2.0& 16.7&  SA& 0.024 &3\\
& & &  & &   & & & & & &\\

19709 & 125845+284133 & 194.6901 & 28.6925 & 0.89$\pm$0.13 & 0.96$\pm$0.13 &35.56$\pm$5.2 & 35$\pm$5.0 & 17.21 & Sc & - &3\\
 &  &  &  & &   &  &  &  & & &\\

19757& Z160-058 & 194.5387 & 28.7086 & 6.48$\pm$0.47 & 8.13 &16.84$\pm$1.21 & 22.0$\pm$5.0&14.28 & Sb&- &1\\
  &  CGCG160058  & &     & &  6.46   & &  16     & & & & 4 \\
& 160-058& & &   & 6.0$\pm$1.0 & &29.0$\pm$3.0 & & &  &5\\
\tableline

\end{tabular}
\end{center}
$^{*}${ Our measured flux is H$\alpha$+[N{\sc ii}] flux as is reported in all references except for Gavazzi et al. (1998) which reported the net H$\alpha$ flux using 5 filters to exclude contribution of [N{\sc ii}] lines from the H$\alpha$ flux measurement. Yagi et al. do not report the errors of the measured H$\alpha$ fluxes and there was no measurement for the EWs }
\\
$^{**}${References: (1) Kennicutt et al. 1984;~ (2) Gavazzi et al. 1998 ;~ (3) Iglesias et al. 2002;~ (4) Gavazzi et al. 2003b (Goldmine);~ (5) Thomas et al. 2008;~ (6) Yagi et al. 2010.}

\end{table}

\footnotesize
\onecolumn
\begin{longtable}[tc]{ c c c c c c c c}
\caption{\normalsize The catalogue of 124 detected H$\alpha$ emitting members of the Coma cluster. First column shows our catalogue ID. Total H$\alpha$+[NII] fluxes and EWs are presented in columns 4 and 5. Column 6 shows the $R$-band magnitudes. All of these sources are confirmed as the Coma cluster members using all available spectroscopic catalogues. Columns 7 and 8 are Hubble type and their redshifts, respectively. The reported Hubble type in column 7 is a combination of the morphological determination of the Coma core members by Michard et al.(2008) and our morphological determination using the SDSS images for the objects in the outlying region of the Coma.}
\label{liscat}

\endfirsthead

\multicolumn{5}{l}{{Continued from previous page}} \\
\tableline
Name & R.A. & Dec.& $H\alpha + [NII]$ flux  & EW$_{H\alpha}$  &  $R$ magnitude & Type & Redshift\\
   &  &  &   &  & & & \\
 (cat ID) & J(2000)& J(2000)& (10$^{-14}$  erg cm$^{-2}$s$^{-1}$)  & ($\AA$) & (mag)& &\\  
\tableline
\tableline

\endhead

\tableline \multicolumn{8}{r}{{Continued on next page}} \\
\endfoot

\tableline 
\endlastfoot

\tableline
Name & R.A. & Dec.& $H\alpha + [NII]$ flux  & EW$_{H\alpha}$  &  $R$ magnitude & Type & Redshift\\
   &  &  &   &  & & & \\
 (cat ID) & J(2000)& J(2000)& (10$^{-14}$  erg cm$^{-2}$s$^{-1}$)  & ($\AA$) & (mag)& &\\  
\tableline
\tableline

6525 &  193.7299 &  27.4127 &  5.12$\pm$0.59 &  3.98$\pm$0.46 &  12.9 &  E &  0.026   \\
8031 &  193.7535 &  27.4834 &  0.42$\pm$0.16 &  4.21$\pm$1.65 &  15.7 &  SBc &  0.023   \\
661 &  193.7713 &  26.7242 &  0.78$\pm$0.19 &  5.92$\pm$1.45 &  15.4 &  E2/3 &  0.022   \\
1251 &  193.8352 &  26.7999 &  0.26$\pm$0.09 &  8.45$\pm$3.08 &  17.0 &  Scd &  0.023   \\
15488 &  193.9447 &  28.1440 &  25.38$\pm$0.40 &  222.95$\pm$3.78 &  15.4 &  SA &  0.020  \\
15912 &  193.9998 &  28.1869 &  0.76$\pm$0.23 &  5.91$\pm$1.76 &  15.4 &  Sb &  0.019   \\
9965$^*$ &  194.0254 &  27.6779 &  2.41$\pm$0.26 &  10.63$\pm$1.13 &  14.9 &  SBa &  0.016   \\
15664 &  194.0457 &  28.1632 &  3.40$\pm$0.26 &  23.87$\pm$1.85 &  15.4 &  S &  0.029   \\
2105$^*$ &  194.1161 &  26.9874 &  3.35$\pm$0.60 &  2.43$\pm$0.44 &  12.9 &  SBab &  0.022   \\
5174$^*$ &  194.1192 &  27.2914 &  5.70$\pm$0.31 &  20.51$\pm$1.11 &  14.6 &  SA &  0.025   \\
2209$^*$ &  194.1214 &  26.9570 &  0.53$\pm$0.39 &  0.87$\pm$0.64 &  13.8 &  SA0 &  0.024   \\
13168 &  194.1242 &  27.9400 &  0.72$\pm$0.37 &  1.93$\pm$0.99 &  14.3 &  E0/1 &  0.022   \\
4556 &  194.1443 &  27.2276 &  1.19$\pm$0.22 &  7.24$\pm$1.34 &  15.2 &  S &  0.024   \\
3161 &  194.1561 &  27.0706 &  0.15$\pm$0.07 &  9.04$\pm$4.08 &  17.7 &  SA &  0.030   \\
4750 &  194.1588 &  27.2178 &  0.11$\pm$0.06 &  10.03$\pm$5.37 &  18.1 &  S &  0.030   \\
16108 &  194.1682 &  28.2177 &  0.38$\pm$0.12 &  12.31$\pm$3.77 &  17.0 &  S &  0.028   \\
3568 &  194.1813 &  27.1788 &  1.64$\pm$0.51 &  1.24$\pm$0.38 &  12.9 &  E3 &  0.025   \\
15981 &  194.1994 &  28.1921 &  0.18$\pm$0.09 &  10.69$\pm$5.07 &  17.7 &  dS &  0.019   \\
3195$^*$ &  194.2069 &  27.0939 &  2.88$\pm$0.31 &  8.74$\pm$0.92 &  14.4 &  SA0/a &  0.023   \\
1750$^*$ &  194.2132 &  26.8989 &  3.82$\pm$0.30 &  9.44$\pm$0.73 &  14.2 &  SA0/a &  0.021   \\
13287 &  194.2214 &  27.9295 &  0.61$\pm$0.26 &  2.88$\pm$1.20 &  14.9 &  S &  0.020   \\
8554 &  194.2681 &  27.5259 &  0.24$\pm$0.15 &  3.07$\pm$1.85 &  16.0 &  SA0 &  0.028   \\
11309 &  194.2690 &  27.7730 &  1.26$\pm$0.25 &  10.19$\pm$2.02 &  15.5 &  S &  0.025   \\
7605 &  194.2892 &  27.4664 &  3.99$\pm$0.33 &  7.76$\pm$0.64 &  13.9 &  E1/3 &  0.025   \\
6814 &  194.2948 &  27.4049 &  0.80$\pm$0.23 &  3.21$\pm$0.93 &  14.7 &  E5 &  0.021   \\
7532 &  194.3515 &  27.4978 &  1.53$\pm$0.70 &  0.89$\pm$0.41 &  12.6 &  E3/5 &  0.025   \\
6928 &  194.3552 &  27.4046 &  0.99$\pm$0.22 &  5.53$\pm$1.26 &  15.1 &  E &  0.016   \\
17778 &  194.3831 &  28.4770 &  11.26$\pm$0.89 &  7.36$\pm$0.58 &  12.8 &  dS0/S &  0.023   \\
8010 &  194.3994 &  27.4932 &  2.24$\pm$0.41 &  3.95$\pm$0.72 &  13.8 &  E0 &  0.024   \\
8064 &  194.4006 &  27.4848 &  0.50$\pm$0.24 &  2.43$\pm$1.16 &  14.9 &  SA0 &  0.024   \\
2850 &  194.4023 &  27.0314 &  1.08$\pm$0.30 &  4.59$\pm$1.26 &  14.8 &  SA/B &  0.025   \\
15756 &  194.4528 &  28.1804 &  1.47$\pm$0.40 &  4.75$\pm$1.29 &  14.5 &  Sb &  0.024   \\
2891 &  194.4861 &  27.0375 &  0.89$\pm$0.24 &  5.73$\pm$1.56 &  15.2 &  SAb &  0.025   \\
7579 &  194.4863 &  27.4426 &  0.35$\pm$0.09 &  13.29$\pm$3.38 &  17.1 &  S &  0.026   \\
14684$^*$ &  194.4907 &  28.0615 &  4.67$\pm$0.33 &  23.90$\pm$1.71 &  15.0 &  merger &  0.027   \\
15946$^*$ &  194.5234 &  28.2425 &  29.92$\pm$0.79 &  35.09$\pm$0.93 &  13.3 &  Sc &  0.024   \\
1190 &  194.5295 &  26.7871 &  0.63$\pm$0.14 &  12.45$\pm$2.86 &  16.5 &  S0 &  0.024   \\
264 &  194.5385 &  26.6643 &  1.19$\pm$0.17 &  17.33$\pm$2.51 &  16.1 &  Scd &  0.024   \\
19757$^*$ &  194.5387 &  28.7086 &  6.48$\pm$0.47 &  16.84$\pm$1.21 &  14.3 &  Sb &  0.025   \\
15352 &  194.5627 &  28.1259 &  1.62$\pm$0.39 &  5.16$\pm$1.23 &  14.5 &  SA0 &  0.025   \\
5711 &  194.5776 &  27.3108 &  5.75$\pm$0.21 &  61.11$\pm$2.31 &  15.7 &  S &  0.025   \\
15526 &  194.5905 &  28.1488 &  1.53$\pm$0.39 &  4.77$\pm$1.21 &  14.5 &  SBa &  0.026   \\
13460 &  194.5910 &  27.9678 &  1.27$\pm$0.49 &  2.32$\pm$0.89 &  13.9 &  SA0 &  0.020   \\
15592 &  194.5923 &  28.1521 &  0.73$\pm$0.28 &  4.32$\pm$1.66 &  15.1 &  SAa &  0.022   \\
15182 &  194.6063 &  28.1289 &  0.84$\pm$0.20 &  10.31$\pm$2.47 &  15.9 &  SBb &  0.027   \\
2913 &  194.6129 &  27.0235 &  0.84$\pm$0.12 &  28.66$\pm$4.04 &  17.0 &  S &  0.025   \\
14065 &  194.6258 &  28.0147 &  1.60$\pm$0.52 &  2.71$\pm$0.88 &  13.8 &  E3 &  0.024   \\
2712 &  194.6289 &  26.9949 &  0.90$\pm$0.10 &  45.78$\pm$5.31 &  17.4 &  Scd &  0.025   \\
14638 &  194.6330 &  28.0496 &  0.64$\pm$0.24 &  5.09$\pm$1.93 &  15.4 &  E/S0 &  0.019   \\
7672 &  194.6336 &  27.4563 &  0.48$\pm$0.18 &  2.28$\pm$0.87 &  14.9 &  SA0 &  0.023   \\
9040$^*$ &  194.6466 &  27.5964 &  5.68$\pm$0.52 &  6.09$\pm$0.56 &  13.3 &  Ep &  0.026   \\
8002 &  194.6467 &  27.4844 &  0.25$\pm$0.03 &  79.86$\pm$10.11 &  19.4 &  dS &  0.021   \\
5001$^*$ &  194.6473 &  27.2647 &  6.21$\pm$0.20 &  36.22$\pm$1.15 &  15.1 &  S &  0.025   \\
15207 &  194.6515 &  28.1137 &  1.00$\pm$0.48 &  2.02$\pm$0.96 &  14.0 &  SAa &  0.023   \\
3913$^*$ &  194.6553 &  27.1766 &  12.30$\pm$0.24 &  51.64$\pm$1.03 &  14.8 &  Sc &  0.026   \\
7599 &  194.6578 &  27.4640 &  0.94$\pm$0.15 &  6.56$\pm$1.07 &  15.3 &  S0/a &  0.021   \\
8707 &  194.6600 &  27.5441 &  0.66$\pm$0.21 &  4.19$\pm$1.34 &  15.2 &  S &  0.020   \\
2846 &  194.6612 &  27.0132 &  0.73$\pm$0.13 &  20.45$\pm$3.51 &  16.8 &  S/ed &  0.023   \\
925 &  194.6660 &  26.7594 &  3.60$\pm$0.20 &  48.49$\pm$2.77 &  16.0 &  Ir/S &  0.025   \\
19709$^*$ &  194.6901 &  28.6925 &  0.89$\pm$0.13 &  35.56$\pm$5.20 &  17.2 &  Sc &  0.022   \\
18782$^*$ &  194.6910 &  28.5430 &  1.75$\pm$0.09 &  101.77$\pm$5.28 &  17.6 &  SA &  0.021   \\
11646 &  194.7030 &  27.8104 &  1.02$\pm$0.38 &  1.77$\pm$0.66 &  13.8 &  SA0 &  0.020   \\
11340 &  194.7171 &  27.7851 &  0.31$\pm$0.28 &  0.95$\pm$0.87 &  14.4 &  SA0 &  0.019   \\
15331 &  194.7227 &  28.1261 &  0.45$\pm$0.30 &  2.35$\pm$1.55 &  15.0 &  Sa &  0.023   \\
13662 &  194.7302 &  27.9647 &  0.45$\pm$0.20 &  2.53$\pm$1.15 &  15.1 &  SA0 &  0.019   \\
11949$^*$ &  194.7332 &  27.8333 &  3.65$\pm$0.37 &  6.86$\pm$0.69 &  13.9 &  Sb &  0.025   \\
11958 &  194.7359 &  27.8220 &  0.17$\pm$0.11 &  3.23$\pm$2.17 &  16.4 &  SA0 &  0.029   \\
9221 &  194.7421 &  27.5947 &  0.80$\pm$0.26 &  3.36$\pm$1.09 &  14.8 &  S0 &  0.020   \\
9870 &  194.7517 &  27.6440 &  0.30$\pm$0.06 &  32.53$\pm$6.20 &  18.3 &  Sm &  0.027   \\
16095 &  194.7575 &  28.2253 &  1.19$\pm$0.36 &  2.08$\pm$0.63 &  13.8 &  Sa &  0.027   \\
15269$^*$ &  194.7586 &  28.1156 &  3.70$\pm$0.35 &  16.70$\pm$1.56 &  14.9 &  SBb &  0.031   \\
15128 &  194.7663 &  28.1237 &  4.98$\pm$0.69 &  5.13$\pm$0.71 &  13.3 &  E3 &  0.026   \\
13664 &  194.7673 &  27.9591 &  0.23$\pm$0.15 &  2.40$\pm$1.58 &  15.8 &  SA0 &  0.023   \\
9508$^*$ &  194.7719 &  27.6443 &  1.48$\pm$0.32 &  3.83$\pm$0.84 &  14.3 &  SBb &  0.018   \\
13913 &  194.7751 &  27.9967 &  1.37$\pm$0.33 &  2.97$\pm$0.72 &  14.1 &  SAa &  0.026   \\
3350 &  194.7786 &  27.0871 &  0.13$\pm$0.04 &  39.45$\pm$12.75 &  19.4 &  S &  0.029   \\
12208$^*$ &  194.7831 &  27.8550 &  0.58$\pm$0.30 &  1.54$\pm$0.80 &  14.3 &  SB0/a &  0.022   \\
14627 &  194.7894 &  28.0409 &  0.22$\pm$0.12 &  3.75$\pm$2.10 &  16.3 &  SBa &  0.024   \\
15695 &  194.7913 &  28.1645 &  0.28$\pm$0.11 &  15.17$\pm$5.88 &  17.5 &  S0/a &  0.029   \\
702 &  194.7924 &  26.7203 &  0.49$\pm$0.11 &  18.64$\pm$4.08 &  17.1 &  Scd &  0.024   \\
9492 &  194.7929 &  27.6199 &  0.65$\pm$0.22 &  3.71$\pm$1.27 &  15.1 &  E2 &  0.019   \\
13883 &  194.8047 &  27.9770 &  3.79$\pm$0.41 &  6.15$\pm$0.66 &  13.7 &  E1 &  0.023   \\
6784 &  194.8071 &  27.4026 &  0.47$\pm$0.25 &  1.10$\pm$0.60 &  14.2 &  E6 &  0.019   \\
14758 &  194.8080 &  28.0763 &  0.59$\pm$0.33 &  1.89$\pm$1.06 &  14.5 &  E1 &  0.026   \\
12899 &  194.8109 &  27.8956 &  0.12$\pm$0.08 &  5.38$\pm$3.50 &  17.3 &  E/S0 &  0.021   \\
13323 &  194.8136 &  27.9707 &  1.62$\pm$0.40 &  2.91$\pm$0.71 &  13.9 &  E5 &  0.016   \\
18249$^*$ &  194.8464 &  28.4886 &  0.82$\pm$0.21 &  6.89$\pm$1.79 &  15.5 &  SA0/a &  0.023   \\
13529 &  194.8722 &  27.9422 &  0.25$\pm$0.06 &  24.29$\pm$5.95 &  18.2 &  Ec &  0.019   \\
14105 &  194.8810 &  28.0466 &  0.71$\pm$0.31 &  2.19$\pm$0.95 &  14.4 &  SA0 &  0.023   \\
16715 &  194.8930 &  28.2771 &  0.14$\pm$0.06 &  14.80$\pm$6.58 &  18.2 &  dE2 &  0.020   \\
12833 &  194.9080 &  27.9073 &  0.55$\pm$0.31 &  1.69$\pm$0.93 &  14.4 &  SA0 &  0.027   \\
19292$^*$ &  194.9174 &  28.6309 &  4.68$\pm$2.56 &  26.4$\pm$1.5 &  15.1 &  Sc &  0.018   \\
13658 &  194.9449 &  27.9739 &  0.55$\pm$0.30 &  1.67$\pm$0.91 &  14.4 &  SB0 &  0.031   \\
16071 &  194.9906 &  28.2467 &  1.06$\pm$0.40 &  1.20$\pm$0.45 &  13.4 &  E0 &  0.022   \\
16373$^*$ &  195.0258 &  28.2514 &  0.30$\pm$0.10 &  6.41$\pm$2.16 &  16.5 &  SBc &  0.021   \\
10061 &  195.0261 &  27.6853 &  0.39$\pm$0.23 &  2.32$\pm$1.36 &  15.2 &  SAa &  0.025   \\
15719 &  195.0380 &  28.1704 &  0.61$\pm$0.27 &  2.31$\pm$1.03 &  14.7 &  SA0/a &  0.023   \\
12433$^*$ &  195.0380 &  27.8664 &  1.59$\pm$0.56 &  3.84$\pm$1.34 &  15.2 &  Irr/S &  0.018   \\
13665 &  195.0688 &  27.9675 &  0.42$\pm$0.26 &  1.74$\pm$1.07 &  14.8 &  SA0/a &  0.015   \\
13356 &  195.0737 &  27.9553 &  2.52$\pm$0.46 &  3.40$\pm$0.62 &  13.6 &  SA0 &  0.023   \\
15548 &  195.0747 &  28.2024 &  8.47$\pm$0.57 &  6.76$\pm$0.45 &  13.0 &  SAa &  0.028   \\
13650 &  195.0754 &  27.9565 &  0.72$\pm$0.29 &  2.42$\pm$0.97 &  14.5 &  SB0 &  0.021   \\
8829 &  195.0796 &  27.5537 &  0.59$\pm$0.28 &  2.29$\pm$1.11 &  14.7 &  S0/E &  0.020   \\
8269 &  195.1398 &  27.5041 &  1.15$\pm$0.18 &  12.56$\pm$1.94 &  15.8 &  S &  0.019   \\
9625$^*$ &  195.1404 &  27.6377 &  5.56$\pm$0.28 &  29.87$\pm$1.50 &  15.1 &  SBc &  0.025   \\
15417 &  195.1482 &  28.1461 &  0.46$\pm$0.28 &  1.68$\pm$1.00 &  14.7 &  SA0/a &  0.018   \\
9036 &  195.1488 &  27.5742 &  0.61$\pm$0.27 &  2.66$\pm$1.18 &  14.9 &  SA &  0.017   \\
19541$^*$ &  195.1548 &  28.6642 &  0.42$\pm$0.13 &  10.10$\pm$3.11 &  16.7 &  SA &  0.024   \\
14516$^*$ &  195.1579 &  28.0575 &  11.90$\pm$0.35 &  38.25$\pm$1.13 &  14.5 &  SAb &  0.026   \\
12977 &  195.1657 &  27.9239 &  1.37$\pm$0.38 &  2.68$\pm$0.74 &  14.0 &  E3 &  0.025   \\
18156 &  195.1694 &  28.5198 &  13.64$\pm$0.31 &  101.59$\pm$2.39 &  15.3 &  S &  0.030   \\
13595 &  195.1782 &  27.9713 &  1.07$\pm$0.40 &  1.85$\pm$0.69 &  13.9 &  SB0/a &  0.021   \\
13353 &  195.1785 &  27.9631 &  0.74$\pm$0.27 &  2.75$\pm$1.02 &  14.6 &  SBa &  0.028   \\
14976 &  195.2027 &  28.0907 &  1.66$\pm$0.41 &  2.68$\pm$0.67 &  13.8 &  E4 &  0.023   \\
15292$^*$ &  195.2034 &  28.1584 &  5.69$\pm$0.52 &  6.09$\pm$0.56 &  13.3 &  Sb &  0.019   \\
14363 &  195.2148 &  28.0429 &  0.84$\pm$0.46 &  1.07$\pm$0.59 &  13.5 &  E5 &  0.029   \\
13887 &  195.2269 &  28.0076 &  1.96$\pm$0.52 &  1.99$\pm$0.53 &  13.3 &  E3 &  0.017   \\
10979$^*$ &  195.2335 &  27.7909 &  18.78$\pm$0.67 &  13.50$\pm$0.48 &  12.9 &  SAb &  0.027   \\
12672$^*$ &  195.2469 &  27.8998 &  0.99$\pm$0.31 &  2.87$\pm$0.90 &  14.4 &  SAa &  0.023   \\
16433 &  195.2880 &  28.2400 &  0.10$\pm$0.04 &  17.72$\pm$6.28 &  18.8 &  dS0 &  0.022   \\
15848 &  195.3038 &  28.1726 &  0.08$\pm$0.05 &  9.93$\pm$6.40 &  18.4 &  dS0 &  0.032   \\
12787 &  195.3068 &  27.9143 &  0.51$\pm$0.19 &  3.96$\pm$1.47 &  15.4 &  E0 &  0.020   \\
15051 &  195.3069 &  28.0831 &  0.14$\pm$0.09 &  4.56$\pm$3.02 &  17.0 &  SA0 &  0.019   \\
18569 &  195.3125 &  28.5218 &  0.83$\pm$0.11 &  28.62$\pm$3.92 &  17.0 &  S &  0.028   \\
\tableline
\end{longtable}

\end{document}